\documentclass[journal,10pt]{IEEEtran}
\usepackage{CJK}
\usepackage{algorithm}
\usepackage{algorithmic}
\usepackage{amsmath}
\usepackage{amssymb}
\usepackage{psfrag}
\usepackage{graphicx}
\usepackage{multirow}
\usepackage{stfloats}
\usepackage{cite}
\usepackage{color}
\usepackage[normalem]{ulem}
\usepackage{arydshln}
\usepackage{enumerate}
\usepackage{bbm}

\newtheorem{mypro}{\textbf{Proposition}}

\begin{document}

\title
{Noncoherent Multiantenna Receivers for Cognitive Backscatter System with Multiple RF Sources}

\author{
Huayan Guo, Qianqian Zhang, Dong Li, and Ying-Chang Liang
%
\thanks{H.Guo, Q.Zhang and Y.-C.Liang are with the Center for Intelligent Networking and Communications, University of Electronic Science and Technology of China (UESTC), Chengdu 611731, China (e-mail: guohuayan@pku.edu.cn).}
\thanks{D.Li is with the Faculty of Information Technology, Macau University of Science and Technology, Macau, China.}
}

\maketitle

\begin{abstract}
Cognitive backscattering, an integration of cognitive radio and backsatter modulation, is emerging as a potential candidate for green \emph{Internet of Things} (IoT). In cognitive backscatter systems, the \emph{backscatter device} (BD) shares not only the same spectrum, but also the same \emph{radio-frequency} (RF) source with the legacy system. In this paper, we investigate the signal transmission problem, in which a basic transmission model is considered which consists of $K$ RF sources, one BD and one reader equipped with $M$ antennas. A non-cooperative scenario is considered, where there is no cooperation between the legacy systems and the backscatter system, and no pilots are transmitted from the RF sources or BD to the reader. The on-off keying differential modulation is adopted to achieve noncoherent transmission. Firstly, through the capacity analyses, we point out that high-throughput backscatter transmission can be achieved when the number of the receive antennas satisfies $M>K$. The \emph{Chernoff Information} (CI) is also derived to predict the detection performance. Next, we address the signal detection problem at the reader. The optimal \emph{soft decision} (SD) and suboptimal \emph{hard decision} (HD) detectors are designed based on the maximum likelihood criterion.
To tackle the non-cooperation challenge, a fully blind channel estimation method is proposed to learn the detection-required parameters based on clustering. Extensive numerical results verify the effectiveness of the proposed detectors and the channel estimation method. In addition, it is illustrated that the increase of $K$ may not necessarily lead to performance degradation when multiple receive antennas are exploited.
\end{abstract}

\begin{IEEEkeywords}
Ambient backscatter, cognitive radio, multiple RF sources, multiple antennas, noncoherent detection, interference cancellation, clustering
\end{IEEEkeywords}


\section{Introduction}
\begin{figure}
\centering
\includegraphics[width=.7\columnwidth]{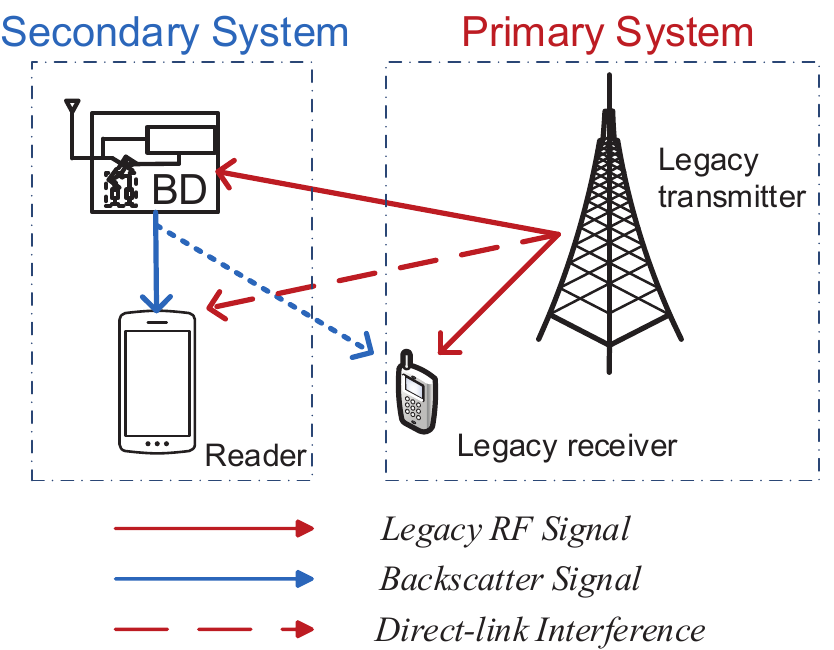}
\caption{A basic CBS illustration.}
\label{Cog_AmBC model}
\vspace{-1.5em}
\end{figure}

\IEEEPARstart{I}{nternet} of Things (IoT) is a key application paradigm for the next generation wireless communication systems.
Due to the energy, cost and complexity constraints, energy- and spectrum efficient communication technologies are desirable for the IoT devices \cite{Stankovic2014DirectionIoT}.
Cognitive backscatter communication is an emerging technology for green IoT to fulfill such demand \cite{LIU2013AMBC}, which is an integration of the well-known cognitive radio concept and the \emph{backscatter communication} (BackCom) technology. To be specific, the backscatter (secondary) system shares not only the same spectrum, but also the same RF source with the legacy (primary) system (e.g., cellular base stations, \emph{digital television} (DTV) transmitters, \emph{wireless fidelity} (Wi-Fi) access points, etc).
A typical \emph{cognitive backscatter system} (CBS) is illustrated in {\figurename~\ref{Cog_AmBC model}}, in which the \emph{backscatter device} (BD) transmits its own information by reflecting the \emph{radio-frequency} (RF) carriers from the legacy  transmitter.
Hence, no power-hungry RF components (e.g., up-converters and power amplifiers) are required, and the power consumption of the BD can be ultra low \cite{Welbourne2009RFID,Boyer2014RFIDBC,ZhangR2013WIPT,Larsson2013SIPT}.
In addition, thanks to the spectrum sharing natural, the CBS also achieves desirable spectrum utilization efficiency \cite{KangX2017RidingPrimary}.

Cognitive backscattering is still in its infant stage, and there are various technical challenges arising from the data transmission and networking perspectives \cite{Niyato2018survey,Niyato2018AmBCMagazine,LiuW2017survey}.
In \cite{LIU2013AMBC}, the first prototype about this conception is introduced which is referred to as the \emph{ambient backscatter communications} (AmBC) for ultra short range communication between two passive tags.
In \cite{Darsena2017ModelandPerformance,darsena2016performance}, early attempts on the information theory and performance analyses are carried out for the \emph{orthogonal frequency division multiplexing} (OFDM) modulated RF source and Gaussian distributed backscatter signals.
In \cite{WangG2018AmBCrate}, a numerical method is presented to calculate the maximum achievable rate of the on-off modulated backscatter signals.
In \cite{Dinesh2015BackFi,Ensworth2017BLE,Talla2017LoRabackscatter}, enormous works are devoted to new prototyping for practical implementation, e.g., ``BakcFi'', ``BLE-Backscatter'', and ``Lora-Backscatter''.
In \cite{lyu2018optimal}, the optimal tradeoff between the energy harvesting and backscatter transmission is investigated by taking the finite battery capacity into account.
In \cite{Niyato2017ImproveNetPerform,Niyato2018TimeSchedulingHTB,wang2018stackelberg,lu2018wireless}, hybrid backscatter network is investigated where the CBS is employed to assist the conventional wireless powered radio networks. In these works, time scheduling protocols are investigated for the tradeoff between the cognitive backscattering and the conventional Backcom techniques.

In this paper, we mainly focus on the receiver design for CBS.
From {\figurename~\ref{Cog_AmBC model}}, the receiver of the backscatter system (i.e., reader) simultaneously receives the backscattered signal from the BD and the \emph{direct-link interference} (DLI) from the legacy system.
Different with the conventional Backcom where the DLI is ``unmodulated'' and can easily be eliminated, the RF carries from the legacy transmitter is ``modulated'' and unknown for the reader in CBS.
The randomness of the unknown strong DLI makes the backscatter symbols very hard to be distinguished.

In \cite{LIU2013AMBC}, noncoherent \emph{energy detector} (ED) is utilized to recover the backscatter symbols.
The performance of the ED and its modified versions are analyzed in \cite{TaoQ2018Manchester,GaoFF2017EDAmBC,wang2016ambient,GaoFF2017NDAmBC}.
In \cite{GaoFF2017NDAmBC,Guo2018IoT}, an interesting error floor problem of ED is pointed out: the \emph{bit error rate} (BER) for backscatter symbol detection converges to a non-zero floor with the increase of the transmitted power at the legacy RF source.
That means, the ED based backscatter system could not benefit from the increase of the RF source transmit power due to the existence of DLI.
As a result, to achieve reliable detection, the transmission rate of the backscatter system becomes quite limited.

To realize the high-speed transmission, it is critical to suppress the DLI at the reader.
Instead of the tag-to-tag communication demonstrated in \cite{LIU2013AMBC}, some works consider a battery-powered reader (employed as a IoT \emph{access point} (AP)), which allows more complicated receiver to remove the DLI effect and to increase the throughput of the backscatter system.
In the literature, some studies suggest to integrate the reader into the legacy transmitter, and then the self-interference cancellation techniques used in full-duplex communications can be exploited to suppress DLI \cite{darsena2016performance,Dinesh2015BackFi}.
Similarly, the reader can also be integrated into the legacy receiver, so that the source signal is jointly decoded together with the backscatter signal \cite{YangG2017CooperativeBC,RZLong2017beamform,DuanRF2017TVTRbistatic}.
Besides the above fully cooperative scenarios, in \cite{YangG2017OFDM}, the OFDM based legacy source is adopted, and the repeating structure of the cyclic prefix is exploited to design the DLI free backscatter waveforms based on some coordination between the legacy transmitter and the BD.
Nevertheless, it is worth noting that the validity of above methods depends on special requirements in terms of transceiver design, RF source modulation, and synchronization between the legacy system and backscatter system; otherwise, these aforementioned methods will fail to work.

\begin{figure}
[!t]
\centering
\includegraphics[width=.6\columnwidth]{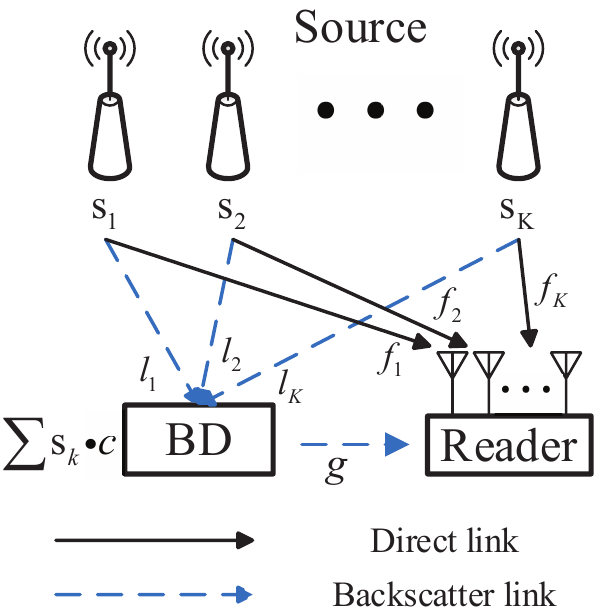}
\caption{The CBS consisting of $K$ RF sources, one BD and one multi-antenna reader. The reader receives signals from both direct link and backscatter link.
}
\vspace{-1.5em}
\label{backscatter model}
\end{figure}

We are interested in developing a general DLI-free detector for various RF sources and application scenarios.
For this purpose, it is assumed that there is no cooperation between the legacy system and the backscatter system.
One early attempt work is presented in \cite{Guo2018IoT}, where only one RF source is considered, and multiple receive antennas are exploited to suppress the DLI.
In this paper, a multiple-RF-source scenario is considered which is illustrated in {\figurename~\ref{backscatter model}}. This assumption is more general and practical. For example, in the smart-home applications, the reader may receive two or even three dominated RF source signals from the neighboring WiFi APs.
In addition, We assume that there is no pilots transmitted from the legacy transmitter or the BD to assist channel estimation at the reader.

The main differences between the CBS shown in {\figurename~\ref{backscatter model}} and the MIMO interference channel are twofold.
First, the backscattering is a multiplicative operation on the incident signals from RF sources in the analog domain.
Therefore, the signal from backscatter link contains a mixture version of the direct-link signals from all the RF sources.
To mitigate the direct link signal straightforwardly may not achieve good performance, which has already been pointed out in \cite{Guo2018IoT}.
Second, the unknown source signals act as unknown channel fading coefficients on the backscatter-link signals.
As a result, the backscatter signal detection is a challenge work, especially without any pilots to assist channel estimation.
In this case, we suggest BD employ on-off differential modulation for noncoherent transmission \cite{LIU2013AMBC,GaoFF2017NDAmBC}.
Novel detection algorithm is required to suppress the DLI from multiple RF sources.

The main works and contributions of this paper are summarized as follows:

\begin{itemize}
\item Firstly, the impacts of multiple RF sources is addressed for the CBS. To be specific, the impacts are twofold: providing carrier emitters for transmission, and causing DLI at the receiver.
    We derive the \emph{Chernoff Information} (CI) \cite{cover2012elements} to analyze the relationship between the detection performance and the system parameters, such as the number of the RF sources and receive antennas.
    Numerical results shown that, the increase of RF sources number may not necessarily lead to performance degradation, when multiple antennas are equipped at the receiver.
\item Secondly, we provide an insight into the error-floor phenomenon through the information-theoretic aspects by deriving the upper bound of the channel capacity.
    It is pointed out that, high-throughput backscatter transmission can be realized by the multi-antenna receiver, as long as the antenna number is larger than the amount of RF sources.
\item Thirdly, the optimal (soft decision) and suboptimal (hard decision) detectors are derived to recover the BD original generated symbols based on the \emph{maximum likelihood} (ML) criterion. Numerical results verify that, the suboptimal detector achieves almost the same performance as that of the optimal one with much lower complexity.
\item Finally, the \emph{channel state information} (CSI) learning problem is addressed for practical implementation, bearing in mind that there is no pilots to be exploited. A novel blind channel estimation method is proposed based on clustering algorithm, thanks to the fact that, the received signals may fall into two clusters corresponding to different backscatter states. A modified version of the relative entropy is designed as a new distance metric to assist clustering. Then, channel estimation is realized by exploiting all the received samples in each cluster.
\end{itemize}

The rest of the paper is organized as follows. Section \ref{system model} outlines the CBS model.
In Section \ref{problem}, we provide some theoretical discussion on the impact of RF Sources, such as the maximum achievable rate and the CI.
Next, in Section \ref{coordinated}, the optimal and suboptimal detectors are derived to recover the original information bits of BD.
Then, for piratical application, blind channel estimation method is proposed in Section \ref{semi-blind}.
Numerical results are provided in Section \ref{simulation} and section \ref{conclusion} concludes the paper.

The notations used in this paper are listed as follows. ${\mathbb E}[\cdot]$ denotes statistical expectation, ${\cal Q} (\cdot)$ is the Q-function, $\mathbbm{1} (\cdot)$ is the indicator function, and ${\rm{sgn}}(\cdot)$ is the signum function. ${\rm Pr}(A)$ denotes the probability of event $A$ happens.
${\rm I}(X;Y)$ denotes the mutual information of random variables $X$ and $Y$, ${\rm H}(X)$ denotes the entropy of $X$, and ${\rm H}(X|Y)$ is the conditional entropy.
${\cal{CN}}(\mu, \sigma^2)$ denotes the \emph{circularly symmetric complex Gaussian} (CSCG) distribution with mean $\mu$ and variance $\sigma^2$.
For any general matrix ${\bf G}$, ${\bf G}^T$ and ${\bf G}^H$ denote the transpose and conjugate transpose, respectively.
${\bf{I}}_{M}$ denotes the $M \times M$ identity matrix.
${\rm{tr}}({\bf S})$ is the trace of a square matrix ${\bf S}$, $|{\bf S}|$ denotes is determinant, ${\rm{rank}}({\bf S})$ denotes its rank, and $\|{\bf S}\|_{\rm F}=\sqrt{{\rm{tr}}({\bf S}{\bf S}^H)}$ denotes its Frobenius norm.
$\|{\bf w}\|$ denotes the Euclidean norm of a vector ${\bf w}$.
The quantity $\max(a,b)$ denotes the maximum between two real numbers $a$ and $b$.
$a \oplus b$ represents the addition modulo $2$.
$|x|$ denotes the absolute value of a complex number $x$, and ${\rm{Re}}(x)$ and ${\rm{Im}}(x)$ denote its real part and imaginary part, respectively.

\section{System Model}\label{system model}

{\figurename~\ref{backscatter model}} depicts the CBS model considered in this paper, which consists of $K$ dominated legacy RF sources, a single-antenna BD and a reader equipped with $M$ antennas.
We only consider one BD transmission in every time slot. When the system contains multiple BDs, they can be scheduled by multiple access control  protocols such as \emph{time division multiple access} (TDMA).
The BD communicates to a neighbouring reader by reflecting the RF signals with different antenna impedances.
The RF carrier wave is dominated by several surrounding RF sources.
For simplicity, flat block fading assumption is adopted, wherein the channel remains constant over consecutive symbol intervals (i.e., a block).

\subsection{RF Source Signals}
Denote $s_{k,n}$ as the $k$-th RF source signal transmitted at time instant $n$. We assume that $s_{k,n}$ is \emph{independent and identically distributed} (i.i.d.) at different time instants, and it follows the standard CSCG distribution, i.e., $s_{k,n} \sim{\cal{CN}}(0,1)$.

The RF source signal received at the reader consists of two components.
One is the signal from the direct link, $r_{m,n}^{\rm{d}}$, which is transmitted directly from the RF sources to the reader.
The other is the signal from the backscatter link, $r_{m,n}^{\rm{b}}$, which is backscattered from the BD to the reader.
Supposing the BD-reader distance is relatively short, the time delay between the receptions of $r_{m,n}^{\rm{d}}$ and $r_{m,n}^{\rm{b}}$ at the reader is negligible \cite{LIU2013AMBC,GaoFF2017NDAmBC,YangG2017CooperativeBC}. The received signals at the $m$-th antenna of the reader can be expressed as
\begin{equation} \label{equ:2_signalx}
y_{m,n}=r_{m,n}^{\rm{d}}+r_{m,n}^{\rm{b}}+u_{m,n},
\end{equation}
where $n=1,2,\ldots,N$, and $u_{m,n}$ is the CSCG noise with zero mean and unit power, i.e., $u_{m,n} \sim{\cal{CN}}(0,1)$.

\subsection{Direct Link}
At time instant $n$, the direct link signal from the $k$-th RF source received at the $m$-th antenna of the reader can be expressed as
\begin{equation} \label{equ:2dl_signalk}
r_{k,m,n}^{\rm{d}}=f_{k,m} \sqrt{P_{{\rm{s}},k}} s_{k,n},
\end{equation}
where $f_{k,m}$ denotes the small-scale fading from the $k$-th RF source to the reader with ${\mathbb E}[|f_{k,m}|^2]=1$ for all $k$ and $m$, and $P_{{\rm{s}},k}$ is the average received power from the $k$-th direct link.
The average received power $P_{{\rm{s}},k}$ is determined by the transmit power $P_{{\rm{t}},k}$ of the $k$-th RF source and the path loss \cite{Rappaport2002WCPP}:
\begin{equation} \label{equ:2PL_direct}
P_{{\rm{s}},k} = \frac{P_{{\rm{t}},k} G_{{\rm{t}},k} G_{\rm{r}} \lambda^2 }{ (4 \pi)^2 L_{{\rm{d}},k}^{\nu_1}},
\end{equation}
where $L_{{\rm{d}},k}$ is the distance between the $k$-th RF source and the reader in meter, $\nu_1$ is the path loss exponent,  $G_{{\rm{t}},k}$ and $G_{\rm{r}}$ are the antenna gain of the $k$-th RF source and the reader, respectively.
Let $\kappa=\frac{{\lambda}^2}{(4 \pi)^2}$, and then \eqref{equ:2PL_direct} becomes
\begin{equation}
P_{{\rm{s}},k} = \frac{\kappa P_{{\rm{t}},k} G_{{\rm{t}},k} G_{\rm{r}}  }{  L_{{\rm{d}},k}^{\nu_1}},
\end{equation}
Then the summation signal $r_{m,n}^{\rm{d}}=\sum_{k=1}^K r_{k,m,n}^{\rm{d}}$ is:
\begin{equation} \label{equ:d_signal}
r_{m,n}^{\rm{d}}=\sum_{k=1}^K f_{k,m} \sqrt{P_{{\rm{s}},k}} s_{k,n}.
\end{equation}

\subsection{Backscatter Link}
The RF source signals received at the BD can be expressed as
\begin{equation} \label{equ:2_signal_st}
s_{n}^{\rm{r}}= \sum_{k=1}^K l_k \sqrt{P_{{\rm{b}},k}} s_{k,n},
\end{equation}
where $l_k$ denotes the small-scale fading from the $k$-th RF source to the BD with ${\mathbb E}[|l_k|^2]$=1, and $P_{{\rm{b}},k}$ is the average available power from the $k$-th RF source before backscattering.
Assuming the same path loss exponent, we have
\begin{equation} \label{equ:2PL_BD}
P_{{\rm{b}},k} = \frac{\kappa P_{{\rm{t}},k} G_{{\rm{t}},k} G_{\rm{b}}  }{  L_{{\rm{b}},k}^{\nu_1}},
\end{equation}
where $L_{{\rm{b}},k}$ is the distance between the $k$-th RF source and the BD, and $G_{\rm{b}}$ is the antenna gain of the BD.

From the antenna scatterer theorem \cite{Fuschini2008BackscatterAnalyze,Griffin2009LinkBudget,Boyer2014RFIDBC}, we assume that BD only has two backscattering states, which are denoted by $c=0$ and $c=1$ (i.e., \emph{on-off keying} (OOK)), respectively.
When $c=0$, only structural mode scattering exists, and when $c=1$, both the structural mode component and antenna mode component are exist.
Since the structural mode scattering is independent to the antenna load impedance and always exists, it has been already contained in the direct link signal. Define $\alpha$ as the reflection coefficient of the antenna mode scattering, and we further suppose that the BD backscattering state $c$ remains unchanged for $N$ consecutive source symbols. Then, during one BD symbol period, the signal backscattered from the BD to the reader is given by
\begin{equation} \label{equ:2_signal_yt}
s_{n}^{\rm{b}}=\alpha s_{n}^{\rm{r}} c,
\end{equation}
where $n=1,2,\ldots,N$, and $0<|\alpha|^2<1$.

The backscatter link signal received at the $m$-th antenna of the reader is expressed as
\begin{equation} \label{equ:2bl_signal}
r_{m,n}^{\rm{b}}= g_m \sqrt{ \frac{ G_{\rm{b}} G_{\rm{r}}\kappa }{ L_{\rm{c}}^{\nu_2} }} s_{n}^{\rm{b}}
,
\end{equation}
where $g_m$ is the small-scale fading from the BD to the reader with ${\mathbb E}[|g_m|^2]$=1,
$L_{\rm{c}}$ is the BD-reader distance, and ${\nu_2}$ is the path loss exponent\footnote{Generally, as $L_{\rm{c}}$ is relatively short, it is a line-of-sight path between the BD and the reader, and we have ${\nu_2}=2$.}.

Substituting \eqref{equ:2PL_direct}, \eqref{equ:2_signal_st} and \eqref{equ:2_signal_yt} into \eqref{equ:2bl_signal}, we have
\begin{equation} \label{equ:2bl_signaln}
\begin{aligned} [b]
r_{m,n}^{\rm{b}}
&=g_m \alpha c \sum_{k=1}^K l_k  \sqrt{ \frac{ \kappa^2  P_{{\rm{t}},k} G_{{\rm{t}},k} G_{\rm{r}} G^2_{\rm{b}}  } { L_{\rm{b}}^{\nu_1} L_{\rm{c}}^{\nu_2}}
     } s_{k,n}\\
&= g_m \alpha c \sum_{k=1}^K l_k  \sqrt{\frac { \kappa P_{{\rm{s}},k} G^2_{\rm{b}} {L_{{\rm{d}},k}^{\nu_1}}} {{L_{{\rm{b}},k}^{\nu_1}} L_{\rm{c}}^{\nu_2}}} s_{k,n}.
\end{aligned}
\end{equation}
Denote the total backscattering power loss of the $k$-th BD as 
\begin{equation}
\widetilde \gamma_k=\frac{\kappa |\alpha|^2  G^2_{\rm{b}}{L_{{\rm{d}},k}^{\nu_1}} }{ {L_{{\rm{b}},k}^{\nu_1}} L_{\rm{c}}^{\nu_2}}.
\end{equation}
Let $\bar{\alpha}=\frac{\alpha}{|\alpha|}$ denote the phase shift due to backscattering.
Then we finally have
\begin{equation}
r_{m,n}^{\rm{b}}= g_m \bar{\alpha} c \sum_{k=1}^K l_k  \sqrt{ \widetilde \gamma_k { P_{{\rm{s}},k} } } s_{k,n}.
\end{equation}

\subsection{Received Signal at The Reader}
Substituting $r_{m,n}^{\rm{d}}$ and $r_{m,n}^{\rm{b}}$ into \eqref{equ:2_signalx}, the received signal at the $m$-th antenna of the reader is expressed as
\begin{equation} \label{equ:2_signal}
y_{m,n}
=\sum_{k=1}^K \sqrt{P_{{\rm{s}},k}} s_{k,n} \left(f_{k,m}+\bar{\alpha} g_m  l_k  \sqrt{ \widetilde \gamma_k } c \right)+u_{m,n}.
\end{equation}
We further denote the average \emph{signal-to-noise ratio} (SNR) of the $k$-th direct link as $\gamma_{{\rm{d}},k} \triangleq P_{{\rm{s}},k}$, and the average SNR of the $k$-th backscatter link as $\gamma_{{\rm{b}},k} \triangleq \widetilde \gamma_k P_{{\rm{s}},k}$.
We assume that the direct link channel response $f_{k,m}$ and the backscatter link channel response $l_k g_m$ are mutually independent.
Then, when the transmit power from one of the RF source increases, or when the distance between the RF sources and the backscatter system ($L_{{\rm{b}},k}$ or $L_{{\rm{d}},k}$) becomes shorter, both $\gamma_{{\rm{d}},k}$ and $\gamma_{{\rm{b}},k}$ increase. Also, when the BD-reader distances become shorter, a larger $\widetilde \gamma_k$ is obtained, resulting in a stronger backscatter link.

With $M$ receive antennas at the reader, we denote the channel response vectors as
\begin{align}
{\bf{f}}_k&=[f_{k,1},f_{k,2},\ldots,f_{k,M}]^T,\\
{\bf{g}}&=[g_1,g_2,\ldots,g_M]^T.
\end{align}
By letting ${\bf{h}}_{k,1}=\sqrt{\gamma_{{\rm{d}},k}} {\bf{f}}_k$ and ${\bf{h}}_{k,2}=\bar{\alpha} l_k \sqrt{ \widetilde \gamma \gamma_{{\rm{d}},k}}{\bf{g}}$, the received signals collected by the $M$ antennas can further be expressed as
\begin{equation} \label{equ:2_signalv}
\begin{aligned} [b]
{\bf{y}}_n
&=\sum_{k=1}^K {{\bf{f}}_k} \sqrt{{\gamma_{{\rm{d}},k}}} s_{k,n}+
{\bf{g}} \sum_{k=1}^K l_k \bar{\alpha} \sqrt{\widetilde \gamma_k {\gamma_{{\rm{d}},k}}} s_{k,n} c
+{\bf{u}}_n\\
&= \sum_{k=1}^K {{{\bf{h}}_{k,1}}} s_{k,n}  +\sum_{k=1}^K {{{\bf{h}}_{k,2}}} s_{k,n} c + {\bf{u}}_n
,
\end{aligned}
\end{equation}
where ${\bf{y}}_n=[y_{1,n},\ldots,y_{M,n}]^T$ and ${\bf{u}}_n=[u_{1,n},\ldots,u_{M,n}]^T$.

\subsection{Frame Structures and Differential Modulation}
\begin{figure}
[!t]
\centering
\includegraphics[width=.99\columnwidth]{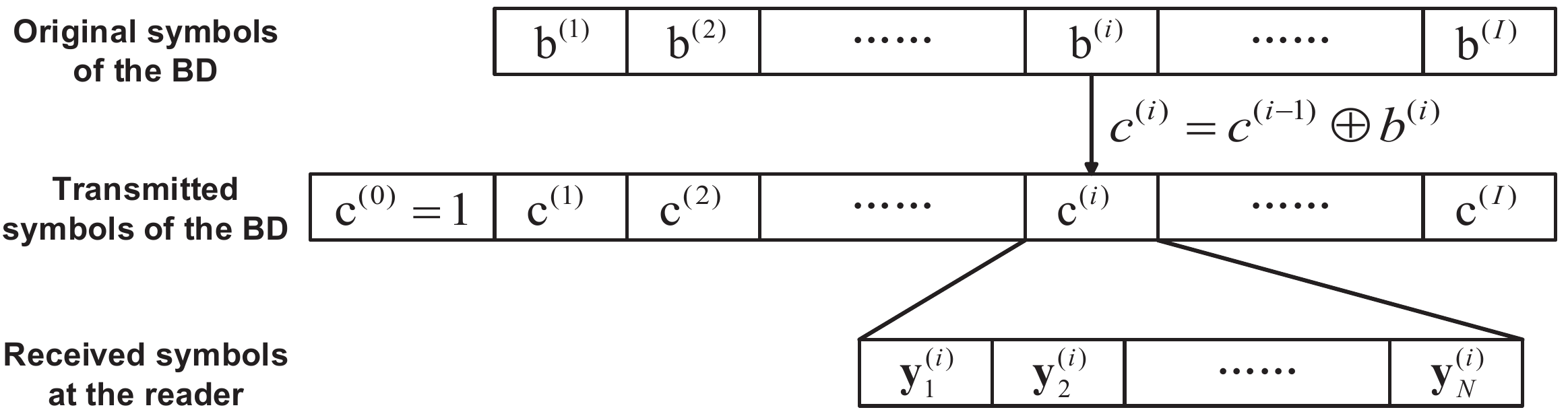}
\caption{Frame structures of involving symbols.
}
\vspace{-1.5em}
\label{c_clusterframe}
\end{figure}

Finally, we adopt a backscatter symbol frame as depicted in {\figurename~\ref{c_clusterframe}}, where each frame consists of $I$ original BD's information bits ${\bf{b}}=[b^{(1)}, b^{(2)},\cdots ,b^{(I)}]$, and $b^{(i)} \in \left\{0,1\right\}$ for all $i=1,2,\cdots,I$. Then the original bits are differentially encoded at the BD as follows
\begin{equation}\label{equ:diffmodulation}
c^{(i)}=c^{(i-1)} \oplus b^{(i)},
\end{equation}
where ${\bf{c}}=[c^{(1)}, c^{(2)},\cdots ,c^{(I)}]$ are the modulated symbols to be transmitted with the reference symbol $c^{(0)}=1$.
According to the block fading assumption, the channels remain invariant during one frame period.
Each BD symbol contains $N$ RF source symbols.
The $n$-th received sample in the $i$-th BD symbol period is
\begin{equation} \label{equ:2_signalvi}
\begin{aligned} [b]
{\bf{y}}_n^{(i)}=\sum_{k=1}^K {\bf{h}}_{k,1} s_{k,n}^{(i)}+ c^{(i)} \sum_{k=1}^K {\bf{h}}_{k,2} s_{k,n}^{(i)} +{\bf{u}}_n^{(i)},
\end{aligned}
\end{equation}
where $n=1,2,\ldots,N$ and $i=0,1,2,\ldots,I$.
Let ${\bf{Y}}^{(i)}=[{\bf{y}}_1^{{(i)}^T} ,{\bf{y}}_2^{{(i)}^T} ,\ldots,{\bf{y}}_N^{{(i)}^T}]^T$ denote the received signal sequence in the $i$-th symbol period.
Our goal is to recover all the $b^{(i)}$ from the observed ${\bf{Y}}^{(i)}$.

\section{The Impact Of the DLI}\label{problem}
In CBS, the relationship between the transmit powers of the RF sources and the achievable rate of the backscatter system is complicated.
On the one hand, the RF sources provide carrier emitters for the BD transmission. Thus, like the traditional BackCom, a strong RF source signal will improve the backscatter transmission.
On the other hand, due to the spectrum sharing, the RF sources simultaneously cause interference to the backscatter system. This is a little like the multiuser MIMO system, where the strong interferences are harmful for the backscatter transmission.
As a result, the CBS model is much different with the conventional communication scenarios.
It is worth to process some analyses to provide an insight about the impact of RF sources on the backscatter transmission.


\subsection{Exploiting Multi-antennas for High-speed Transmission}
In CBS detection, a well-known phenomenon for the single-antenna receiver is the ``error floor'': the BER for backscatter symbol detection converges to a non-zero floor even though the transmit power of the RF sources increases to infinite \cite{GaoFF2017NDAmBC,Guo2018IoT}.
Due to that, given a BER requirement, the transmission rate is limited.
In this subsection, we try to provide an explanation on the rate limitation of single-antenna receiver from the information theoretic perspective.
Then, we will show that to exploit multiple receive antennas can realize high-speed backscatter transmission.
For simplicity, we set $N=1$, and the efficiency loss of the differential modulation is also ignored as well.

\subsubsection{Single-antenna Receiver}
When there is only one RF source and the receiver is equipped with single antenna, the received signal at the reader is
\begin{equation} \label{equ:2_oneantenna}
{{y}} =l g \sqrt{\widetilde \gamma {\gamma_{\rm{d}}}} s c +f \sqrt{{\gamma_{\rm{d}}}} s+u.
\end{equation}
The maximum achievable rate is $R_{\{M=1\}}={\max_{p(c)}}{\rm I}(c;{{y}})$.
As shown in \eqref{equ:2_oneantenna}, the backscatter symbol $c$ is corrupted by the unknown source signal $s$.
Due to the unknown $s$, ${R}_{\{M=1\}}$ is hard to be derived.

We consider a scenario, where the BD is allowed to exploit $s$  to design the transmit codewords. Then \eqref{equ:2_oneantenna} becomes
\begin{equation} \label{equ:2_oneanew}
y =l g \sqrt{\widetilde \gamma {\gamma_{\rm{d}}}} \widetilde{{c}} +f \sqrt{{\gamma_{\rm{d}}}} s+u,
\end{equation}
where $\widetilde{{c}}$ is the new transmitted symbol by BD which drops the interference of $s$ to achieve the optimal rate.
Denote the achievable rate in this scenario as ${\widetilde{R}}_{\{M=1\}}$, and we have
\begin{equation} \label{equ:R_oneantenna}
{\widetilde{R}}_{\{M=1\}}
=  \log_2 (1+\frac{\widetilde \gamma \gamma_{\rm{d}}|l g|^2}{1+\gamma_{\rm{d}}|f|^2})
.
\end{equation}
Obviously, ${\widetilde{R}}_{\{M=1\}}$ is an upper bound of ${R}_{\{M=1\}}$ (i.e., ${R}_{\{M=1\}}\leq {\widetilde{R}}_{\{M=1\}}$).
Then we have following proposition:

\begin{mypro}\label{pro1}
When there is only one RF source and the receiver is equipped with single antenna, the achievable rate converges to a finite value  with the increase of $\gamma_{\rm{d}}$.
\end{mypro}

\begin{IEEEproof}
Let ${\widetilde{R}}_{\{M=1\}}^{\infty}={\lim_{\gamma_{\rm{d}} \rightarrow \infty}} {\widetilde{R}}_{\{M=1\}}$ denote the extreme rate.
Then according to \eqref{equ:R_oneantenna}, we obtain
\begin{equation} \label{equ:Rup_oneantenna}
\begin{aligned} [b]
{\widetilde{R}}_{\{M=1\}}^{\infty}&= \lim_{\gamma_{\rm{d}}\rightarrow +\infty} \log_2 (1+\frac{\widetilde \gamma \gamma_{\rm{d}}|l g|^2}{1+\gamma_{\rm{d}}|f|^2})\\
&= \log_2 (1+\frac{\widetilde \gamma |l g|^2}{|f|^2})
.
\end{aligned}
\end{equation}
Therefore, ${\widetilde{R}}_{\{M=1\}}^{\infty}$ is dominated by the relative SNR $\widetilde \gamma$, and obviously, ${\widetilde{R}}_{\{M=1\}}^{\infty}<+\infty$.
\end{IEEEproof}

In CBS, $\widetilde \gamma$ is very small (less than $-20$ dB), and ${\widetilde{R}}_{\{M=1\}}^{\infty}$ is quiet limited due to the existence of the DLI. That is the key reason of the error floor phenomenon.
As a result, to suppress the DLI is the critial task to realize high-speed backscatter transmission.

\subsubsection{Multi-antenna Receiver}
To realize high-speed backscatter transmission, we exploit the receiving diversity in the spatial domain with the multi-antenna receiver ($M\geq 2$). Suppose that there are $K$ RF sources, according to \eqref{equ:2_signalv}, the received signal at the receiver is
\begin{equation} \label{equ:Mantenna}
{\bf{y}}
={\bf{g}} \sum_{k=1}^K l_k \sqrt{\widetilde \gamma_k {\gamma_{{\rm{d}},k}}} s_k c +\sum_{k=1}^K {{\bf{f}}_k} \sqrt{{\gamma_{{\rm{d}},k}}} s_k+{\bf{u}}.
\end{equation}
The theoretical maximum achievable rate is denoted as $R_{\{M,K\}}={\max_{p(c)}}{\rm I}(c;{\bf{y}})$.

As the $K$ RF sources are mutually independent, we assume ${{\bf{f}}_1}$, ${{\bf{f}}_2}$, $\cdots$, ${{\bf{f}}_K}$ and ${\bf{g}}$ are linearly independent, i.e., the equation $a_1 {{\bf{f}}_1}+a_2 {{\bf{f}}_2}+\cdots+a_K {{\bf{f}}_K}+a_{K+1} {\bf{g}}={\bf{0}}$
is only satisfied by $a_1=a_2=\cdots=a_{K+1}=0$.
Then we have following proposition:

\begin{mypro}\label{pro2}
When the receive antenna number is larger than $K$, i.e., $M>K$, infinite $R_{\{M,K\}}$ can be achieved as long as the transmit power of one  RF source (${\gamma_{{\rm{d}},k_0}}$) approaches infinity.
\end{mypro}

\begin{IEEEproof}
See Appendix \ref{M_larger_K}.
\end{IEEEproof}

\subsection{Chernoff Information for the On-off Modulation}
To achieve low implementation cost, the BD usually only has two backscatter states, i.e., $c$ employs the on-off modulation.
However, the close-form expression of the maximum achievable rate in this scenario is unavailable (A numerical method is shown in Appendix \ref{A_onoff_rate}).
In this subsection, we resort to the CI as a tractable metric to predict the optimal detection performance, which is the maximum achievable error exponent \cite[eq 11.230]{cover2012elements}:
\begin{equation} \label{equ:chernoff}
D= \max \lim_{N\rightarrow \infty} -\frac{1}{N} \ln {P_{\rm e}},
\end{equation}
where $P_{\rm e}$ is the decision BER.

According to \cite[eq 11.239]{cover2012elements}, the standard definition of CI is
\begin{equation} \label{equ:Dchernoff}
D= - \min_{0\leq u \leq 1} \ln\left( {p^{u}({\bf{y}} | c=1 )} {p^{1-u}({\bf{y}} | c=0 )} \right),
\end{equation}
where ${p({\bf{y}} | c=1 )}$ and ${p({\bf{y}} | c=0 )}$ are the conditional \emph{probability density functions} (PDFs) given $c$.
Since both the RF source signals $s_k$ and the noise $\bf u$ follow CSCG distribution, we have:
\begin{align}
{p({\bf{y}} | c=1 )}
&= \frac{1}{\pi^M |{\bf{C}}_1|} e^{
        - {{\bf{y}}}^H {\bf{C}}_1^{-1} {\bf{y}}
} \label{equ:Gpdfc1}
,\\
{p({\bf{y}} | c=0 )}
&= \frac{1}{\pi^M |{\bf{C}}_0|} e^{
        - {{\bf{y}}}^H {\bf{C}}_0^{-1} {\bf{y}}
}, \label{equ:Gpdfc0}
\end{align}
where ${\bf{C}}_1$ and ${\bf{C}}_0$ are the channel statistical covariance matrices:
\begin{align}
{\bf{C}}_1&=\sum_{k=1}^K ({{\bf{h}}_{1,k}+{\bf{h}}_{2,k}}) ({{\bf{h}}_{1,k}+{\bf{h}}_{2,k}})^H+ {\bf{I}}_{M}. \label{equ:C1}\\
{\bf{C}}_0&=\sum_{k=1}^K {{\bf{h}}_{1,k}} {{\bf{h}}_{1,k}^H}+ {\bf{I}}_{M}, \label{equ:C0}
\end{align}

Let $G(u)={p^{u}({\bf{y}} | c=1 )} {p^{1-u}({\bf{y}} | c=0 )}$, where $u \in [0,1]$.
According to \eqref{equ:Dchernoff}, the CI is $D= - \min_{0\leq u \leq 1} \ln\left( G(u) \right)$.
Substituting \eqref{equ:Gpdfc1} and \eqref{equ:Gpdfc0} into $G(u)$, we have
\begin{equation} \label{equ:funcG}
\ln G(u)= \ln\left(\left|{\bf{K}}(u)\right|\right)-u \ln (|{\bf{C}}_1|)-(1-u) \ln (|{\bf{C}}_0|),
\end{equation}
where
\begin{equation}
{\bf{K}}^{-1}(u)=u {\bf{C}}_1^{-1}+(1-u) {\bf{C}}_0^{-1}.
\end{equation}
Based on the matrix differentiation identities:
\begin{align}
\frac{\rm d}{{\rm d}u} \ln\left(\left|{\bf{K}}(u)\right|\right)&={\rm tr} \left[{\bf{K}}^{-1}(u) \frac{{\rm d}{\bf{K}}}{{\rm d}u} (u)\right],
\\
\frac{\rm d}{{\rm d}u} {\bf{K}}^{-1}(u)&=-{\bf{K}}^{-1}(u) \frac{{\rm d}{\bf{K}}}{{\rm d}u} (u) {\bf{K}}^{-1}(u),
\end{align}
we have:
\begin{equation}
\frac{\rm d}{{\rm d}u} \ln\left(G(u)\right)=-{\rm tr} \left({\bf{K}}(u)\left({\bf{C}}_1^{-1}-{\bf{C}}_0^{-1}\right)\right)-\ln{\frac{|{\bf{C}}_1|}{|{\bf{C}}_0|}}.
\end{equation}
Solving $\frac{\rm d}{{\rm d}u} \ln\left(G(u^\ast)\right)=0$, CI is obtained
\begin{equation}
D= - \ln\left( G(u^\ast) \right).
\end{equation}
Further discussions will be presented in Section \ref{OOK_rate_sim}, which illustrates the impact of $K$ and $M$ on the CI.

\section{Maximum Likelihood Detectors}\label{coordinated}
In this section, we investigate on the backscatter signal detection for the multi-antenna receiver based on the ML criterion.
Based on \emph{\textbf{Proposition}} \ref{pro2}, we assume that $M>K$ to achieve reliable performance.
The backscatter system spanning $N$ consecutive source symbols to make sure that the BD transmission causes no negative impact on the legacy system \cite{YangG2017CooperativeBC,RZLong2017beamform}.



\subsection{Soft Messages of ${\bf{c}}$}
From the frame structures shown in {\figurename~\ref{c_clusterframe}}, the symbol vector ${\bf{c}}$ is the sufficient statistic of the original signal vector ${\bf{b}}$.
Hence, recovering ${\bf{c}}$ is the first step to decode ${\bf{b}}$.

During one symbol period, $c^{(i)}$ is equal to ``$0$'' or ``$1$''. As a result, to recover $c^{(i)}$ is equivalent to distinguish whether the backscatter-link signal is present or absent:
\begin{equation} \label{equ:3_sensingh}
{\bf{y}}_n^{(i)}=
\begin{cases}
\sum_{k=1}^K {\bf{h}}_{k,1} s_{k,n}^{(i)}+{\bf{u}}_n^{(i)}, & {\rm{if}} ~~ c^{(i)} = 0,
\\
\sum_{k=1}^K ({\bf{h}}_{k,1}+{\bf{h}}_{k,2}) s_{k,n}^{(i)}+{\bf{u}}_n^{(i)}, & {\rm{if}} ~~ c^{(i)} = 1.
\end{cases}
\end{equation}
where $n=1,2,\ldots,N$, and $i=1,2,\ldots,I$.

As we know, ${\bf{y}}_n^{(i)}$ is a CSCG distributed vector with conditional PDFs given $c^{(i)}$:
\begin{align}
{p({\bf{y}}_n^{(i)} | c^{(i)}=0 )}
&= \frac{1}{\pi^M |{\bf{C}}_0|} e^{
        - {{\bf{y}}_n^{(i)}}^H {\bf{C}}_0^{-1} {\bf{y}}_n^{(i)}
}, \label{equ:3_Gpdfc0}\\
{p({\bf{y}}_n^{(i)} | c^{(i)}=1 )}
&= \frac{1}{\pi^M |{\bf{C}}_1|} e^{
        - {{\bf{y}}_n^{(i)}}^H {\bf{C}}_1^{-1} {\bf{y}}_n^{(i)}
} \label{equ:3_Gpdfc1}
.
\end{align}
Then the likelihood function of the received signal sequence ${\bf{Y}}^{(i)}=[{\bf{y}}_1^{{(i)}^T} ,{\bf{y}}_2^{{(i)}^T} ,\ldots,{\bf{y}}_N^{{(i)}^T}]^T$ is
\begin{equation} \label{equ:LFY}
{\mathcal{L}}{({\bf{Y}}^{(i)} | c^{(i)} )}= \prod_{n=1}^N {p({\bf{y}}_n^{(i)} | c^{(i)} )}.
\end{equation}
Substituting \eqref{equ:3_Gpdfc0} and \eqref{equ:3_Gpdfc1} into \eqref{equ:LFY}, we have
\begin{align}
{\mathcal{L}}{({\bf{Y}}^{(i)} | c^{(i)}=0 )}
&= \prod_{n=1}^N {
\frac{1}{\pi^M |{\bf{C}}_0|} e^{
        - {{\bf{y}}_n^{(i)}}^H {\bf{C}}_0^{-1} {\bf{y}}_n^{(i)}
}} \label{equ:YLH0}
,\\
{\mathcal{L}}{({\bf{Y}}^{(i)} | c^{(i)}=1 )}
&= \prod_{n=1}^N{
\frac{1}{\pi^M |{\bf{C}}_1|} e^{
        - {{\bf{y}}_n^{(i)}}^H {\bf{C}}_1^{-1} {\bf{y}}_n^{(i)}
}} \label{equ:YLH1}
.
\end{align}
The likelihood functions \eqref{equ:YLH0} and \eqref{equ:YLH1} contain the whole information of $c^{(i)}$ inferred from ${\bf{Y}}^{(i)}$, which is referred to as the ``soft message'' of $c^{(i)}$.

\subsection{ML Detector for the Original Symbols ${\bf{b}}$}\label{ML_detection}
From \eqref{equ:diffmodulation}, the BD encodes the original symbol $b^{(i)}$ via differential modulation.
Therefore, $c^{(i)}$ and $c^{(i-1)}$ are the sufficient statistics of $b^{(i)}$. In the following, we will design the ML detectors for $b^{(i)}$ based on the soft messages of $c^{(i)}$ and $c^{(i-1)}$, respectively.

\subsubsection{Soft Decision}
Define ${\bf{z}}^{(i)}=[{\bf{Y}}^{(i-1)};{\bf{Y}}^{(i)}]$.
According to \eqref{equ:diffmodulation}, the likelihood functions of ${\bf{z}}^{(i)}$ conditional on $b^{(i)}$ are expressed as follows based on the soft messages of $c^{(i)}$ and $c^{(i-1)}$:
\begin{equation} \label{equ:LFYb0}
\begin{aligned} [b]
{\mathcal{L}}{({\bf{z}}^{(i)} | b^{(i)}=0 )}&= {\mathcal{L}}{({\bf{z}}^{(i)} | c^{(i-1)}=c^{(i)} )}
\\
&={\mathcal{L}}{({\bf{Y}}^{(i-1)} | c^{(i-1)}=0 )} {\mathcal{L}}{({\bf{Y}}^{(i)} | c^{(i)}=0 )}\\
&\quad +{\mathcal{L}}{({\bf{Y}}^{(i-1)} | c^{(i-1)}=1 )} {\mathcal{L}}{({\bf{Y}}^{(i)} | c^{(i)}=1 )}
,
\end{aligned}
\end{equation}
\begin{equation} \label{equ:LFYb1}
\begin{aligned} [b]
{\mathcal{L}}{({\bf{z}}^{(i)} | b^{(i)}=1 )}&= {\mathcal{L}}{({\bf{z}}^{(i)} | c^{(i-1)}\neq c^{(i)} )}
\\
&={\mathcal{L}}{({\bf{Y}}^{(i-1)} | c^{(i-1)}=0 )} {\mathcal{L}}{({\bf{Y}}^{(i)} | c^{(i)}=1 )}\\
&\quad +{\mathcal{L}}{({\bf{Y}}^{(i-1)} | c^{(i-1)}=1 )} {\mathcal{L}}{({\bf{Y}}^{(i)} | c^{(i)}=0 )}
.
\end{aligned}
\end{equation}
Then, the ML detector for the original symbols $b^{(i)}$ is expressed as follows:
\begin{equation}\label{equ:soft}
{\mathcal{L}}{({\bf{z}}^{(i)} | b^{(i)}=0 )}
\mathop{\gtrless}_{{\hat b}^{(i)}=1}^{{\hat b}^{(i)}=0}
{\mathcal{L}}{({\bf{z}}^{(i)} | b^{(i)}=1 )}
,
\end{equation}
where ${\hat b}^{(i)}$ is the decision result.
The ML detector in \eqref{equ:soft} is usually referred to as ``\emph{soft decision}'' (SD) method.

\subsubsection{Hard Decision}
In some scenarios, a two-step ML decision method are applied instead of the SD method in \eqref{equ:soft} for simplicity.

In the first step, the ``\emph{hard decision}'' (HD) of $c^{(i)}$ is obtained based on the ML criterion:
\begin{equation}\label{equ:hardmc}
{\mathcal{L}}{({\bf{Y}}^{(i)} | c^{(i)}=0 )}
\mathop{\gtrless}_{{\hat c}^{(i)}=1}^{{\hat c}^{(i)}=0}
{\mathcal{L}}{({\bf{Y}}^{(i)} | c^{(i)}=1 )}
,
\end{equation}
Substituting \eqref{equ:YLH0} and \eqref{equ:YLH1} into \eqref{equ:hardmc}, we obtain the ML detector of the transmitted symbol $c^{(i)}$ as follows:
\begin{equation}\label{equ:3_opt_CSCGv}
\sum_{n=1}^N
{{\bf{y}}_n^{(i)}}^H \left({\bf{C}}_0^{-1}-{\bf{C}}_1^{-1}\right) {\bf{y}}_n^{(i)}
\mathop{\gtrless}_{{\hat c}^{(i)}=0}^{{\hat c}^{(i)}=1}
N \ln \frac {|{\bf{C}}_1|} {|{\bf{C}}_0|}
,
\end{equation}
where $[{\hat c}^{(1)}, {\hat c}^{(2)},\cdots ,{\hat c}^{(I)}]$ are the decision results of $\bf c$.

In the second step, the decision result ${\hat b}^{(i)}$ is derived based on the differential modulation relationship shown in \eqref{equ:diffmodulation}:
\begin{equation} \label{equ:hard}
{\hat b}^{(i)}={\hat c}^{(i-1)} \oplus {\hat c}^{(i)}.
\end{equation}

When $M=1$, the ML detector for $c^{(i)}$ in \eqref{equ:3_opt_CSCGv} is equivalent to the ED, and thus the above HD method is also known as the joint-ED in \cite{GaoFF2017NDAmBC}.

\section{Clustering: Blind Channel Estimation for Practical Implementation}\label{semi-blind}
In section \ref{coordinated}, we have designed the optimal SD and suboptimal HD detectors with the knowledge of CSI (i.e., ${\bf{C}}_0$ and ${\bf{C}}_1$).
In conventional communication system, the CSI is usually estimated with the aid of pilot symbols.
Nonetheless, in CBS, the backscatter link is very weak ($\widetilde \gamma$ is generally smaller than $-20$ dB), so that a long pilot sequence is need which may incur severer efficiency loss \cite{GaoFF2017NDAmBC}.
Thus, channel estimation is another critical challenge for practical implementation, especially without the cooperation with the legacy system.
In this section, we try to propose a fully blind channel estimation method, which do no need any pilot symbols to achieve high transmission efficiency.

Based on that the BD transmits binary modulated signals, we suggest a fully blind channel estimation method via clustering \cite{Hastie2009slearn}.
Our intuition is summarized below:
\begin{itemize}
\item According to \eqref{equ:3_sensingh}, the received signals $\left\{{\bf{Y}}^{(i)}\right\}^I_{i=1}$ can be grouped into two clusters since they come from two different distributions (see equations \eqref{equ:YLH0} and \eqref{equ:YLH1}).
\item Thanks to the differential modulation, we actually do not need to map the two groups to the transmission states ``$c=0$'' and ``$c=1$''.
The CSI for each group can be carried out by combining all the received samples belonging to the same cluster.
\end{itemize}
The details of the clustering method will be presented in the next three subsections.

\subsection{Feature Extraction}\label{semi-blind-feature}
For a single antenna receiver, the energy of received symbol is a good feature to group all the $\left\{{\bf{Y}}^{(i)}\right\}^I_{i=1}$ into two clusters. Based on that, clustering is realized through a simple sorting algorithm, in which the first half and the second half of energy levels are classified into two groups, respectively \cite{{GaoFF2017NDAmBC}}.
However, when the reader has multiple antennas, this is a challenge task due to the high dimension of ${\bf{Y}}^{(i)}$.
To extract a proper feature from ${\bf{Y}}^{(i)}$ is the first key task for clustering.

Denote the channel statistical covariance matrix corresponding to ${\bf{Y}}^{(i)}$ as ${\bf{C}}^{(i)}$.
Then we have
\begin{equation}
{\bf{C}}^{(i)}=
\begin{cases}
{\bf{C}}_0, & {\rm{if}} ~~ c^{(i)} = 0,
\\
{\bf{C}}_1, & {\rm{if}} ~~ c^{(i)} = 1.
\end{cases}
\end{equation}
The ML estimation of ${\bf{C}}^{(i)}$ is the sample covariance matrix:
\begin{equation} \label{equ:samplecm}
{{\bf{R}}}^{(i)}=\frac{1}{N} \sum_{n=1}^{N}
{
    {\bf{y}}_n^{(i)} {{\bf{y}}_n^{(i)}}^H
}
,
\end{equation}
which contains all the useful information of ${\bf{Y}}^{(i)}$.
In most cases, ${{\bf{R}}}^{(i)}$ is accurate enough.
Nevertheless, the estimation accuracy can be further improved by exploiting the spatial sparsity of ${\bf{C}}^{(i)}$, which will be significant when the antenna number $M$ is much larger than $K$.

Let us divide ${\bf{C}}^{(i)}$ into two components: the signal component ${{\bf{C}}}_{s}$ and the noise component ${{\bf{C}}}_u$. Since the noise follows the standard CSCG distribution, ${{\bf{C}}}_u={\bf{I}}_{M}$. Then we have
\begin{equation}\label{equ:estC}
{\bf{C}}^{(i)}={{\bf{C}}}_{s}+{\bf{I}}_{M}
.
\end{equation}
For different transmit symbols ($c=0$ or $c=1$), ${{\bf{C}}}_{s}$ is expressed as follows
\begin{align}
{\bf{C}}_{s,0}&=\sum_{k=1}^K {{\bf{h}}_{1,k}} {{\bf{h}}_{1,k}^H}, \label{equ:Cs0} \\
{\bf{C}}_{s,1}&=\sum_{k=1}^K ({{\bf{h}}_{1,k}+{\bf{h}}_{2,k}}) ({{\bf{h}}_{1,k}+{\bf{h}}_{2,k}})^H. \label{equ:Cs1}
\end{align}
Thus ${\rm{rank}}({{\bf{C}}}_{s})=K$, and there are some redundant elements in ${{\bf{C}}}_{s}$ when the the number of the dominated RF sources is smaller than $M-1$.
Consequently, a better estimation of ${{\bf{C}}}_{s}$ is:
\begin{equation}\label{equ:clusterRF}
{{\bf{R}}}^{(i)}_{s}= \sum_{k=1}^K ({\rm \lambda}_{k}-1) {\bf{w}}_{k} {\bf{w}}_{k}^H
,
\end{equation}
where ${\rm \lambda}_{k}$ is the $k$-th largest eigenvalue of ${{\bf{R}}}^{(i)}$, and ${\bf{w}}_{k}$ is the corresponding eigenvector.
Combining \eqref{equ:estC} and \eqref{equ:clusterRF}, we have a better estimation of the channel covariance matrix, which will be employed as the features for clustering:
\begin{equation}\label{equ:improvedesR}
{\hat{\bf{R}}}^{(i)}= {{\bf{R}}}^{(i)}_{s}+{\bf{I}}_{M}.
\end{equation}

\subsection{Distance Design}
After feature extraction, the second key task for clustering is to measure the difference between features ${\hat{\bf{R}}}^{(i)}$ and ${\hat{\bf{R}}}^{(j)}$, which is referred to as the ``distance'' in the statistical machine learning.

Based on ${\hat{\bf{R}}}^{(i)}$, we define the pseudo-PDF as follows
\begin{equation}\label{equ:pPDF}
{\hat{p}}{({\bf{Y}} | {\hat{\bf{R}}}^{(i)} )}
= \prod_{n=1}^N {
\frac{1}{\pi^M |{\hat{\bf{R}}}^{(i)}|} e^{
        - {\bf{y}}_n^H ({{\hat{\bf{R}}}^{(i)}})^{-1} {\bf{y}}_n
}},
\end{equation}
where ${\bf{Y}}=[{\bf{y}}_1^T,{\bf{y}}_2^T,\ldots,{\bf{y}}_n^T]^T$ denotes the received signal sequence.
It is known that, the relative entropy between two $M$-dimension zero-mean multivariate CSCG distributions with covariance matrices ${\bf \Sigma}_0$ and ${\bf \Sigma}_1$ is
\begin{equation}\label{equ:KL}
D_{\rm KL} ({\bf \Sigma}_0 \parallel {\bf \Sigma}_1)=\frac{1}{2}\left(
    {\rm{tr}}\left({\bf \Sigma}_1^{-1} {\bf \Sigma}_0\right)-M+\ln \frac {|{\bf \Sigma}_1|} {|{\bf \Sigma}_0|}
\right)
.
\end{equation}

Based on \eqref{equ:KL}, we propose a new distance for clustering, which is defined by
\begin{equation}\label{equ:distance}
\begin{aligned} [b]
d_{ij}&= \left(D_{\rm KL} ({\hat{\bf{R}}}^{(i)} \parallel {\hat{\bf{R}}}^{(j)})+D_{\rm KL} ({\hat{\bf{R}}}^{(j)} \parallel {\hat{\bf{R}}}^{(i)})\right)\\
&=\frac{1}{2}{\rm{tr}}\left(({\hat{\bf{R}}}^{(i)})^{-1} {\hat{\bf{R}}}^{(j)}+({\hat{\bf{R}}}^{(j)})^{-1} {\hat{\bf{R}}}^{(i)}\right)-M
.
\end{aligned}
\end{equation}
Obviously, the distance in \eqref{equ:distance} satisfies $d_{ii}=0$ and $d_{ij}=d_{ji}$.

\subsection{Clustering}
Based on the feature ${\hat{\bf{R}}}^{(i)}$ and the distance metric $d_{ij}$, we present the clustering algorithm.
For clustering, the key task is to find the cluster centroids according to a given distance metric.
We suggest a two-step centroid search method based on the distance defined in \eqref{equ:distance}:
\begin{itemize}
\item Firstly, we randomly choose one element ${\hat{\bf{R}}}^{(i)}$, then the first cluster centroid is ${\hat{\bf{R}}}^{(a_0)}$ where $a_0= \arg \max_{j} d_{ij}$.
\item Similarly, the second cluster centroid is ${\hat{\bf{R}}}^{(a_1)}$ where $a_1= \arg \max_{j} d_{a_0 j}$.
\end{itemize}

Using ${\hat c}^{(i)}=0$ to indicate that ${\hat{\bf{R}}}^{(i)}$ belongs to the first cluster, and  ${\hat c}^{(i)}=1$ to indicate that ${\hat{\bf{R}}}^{(i)}$ belongs to the second cluster, the clustering is implemented as follow:
\begin{equation}\label{equ:clustering}
d_{a_0 i}
\mathop{\gtrless}_{{\hat c}^{(i)}=0}^{{\hat c}^{(i)}=1}
d_{a_1 i}
.
\end{equation}
Based on the clustering results, all the received symbols in the same cluster are combined to realize the estimation of the channel statistical covariance matrices:
\begin{align}
\hat{\bf{C}}_0 &=\frac{1}{\sum_{i=1}^I (1-{\hat c}^{(i)})} \sum_{i=1}^I (1-{\hat c}^{(i)}) \hat{\bf{R}}^{(i)}, \label{equ:clusterR0}\\
\hat{\bf{C}}_1 &=\frac{1}{\sum_{i=1}^I {\hat c}^{(i)}} \sum_{i=1}^I {\hat c}^{(i)} \hat{\bf{R}}^{(i)}. \label{equ:clusterR1}
\end{align}
Subsequently, signal detection can be executed based on the SD or HD in Section \ref{coordinated}.

\section{Numerical Results}\label{simulation}
In this section, we resort to numerical examples to evaluate the multi-antenna receiver schemes.
All the channels are assumed experience independent Rayleigh fading, and for simplicity, a homogenous scenario is assumed in which all the $\widetilde \gamma_k$ and $\gamma_{{\rm{d}},k}$ corresponding to different RF sources are the same.
Denote the summation of the incident direct-link power by $\gamma_{{\rm{d}}}=\sum \gamma_{{\rm{d}},k}$, and the relative SNR by $\widetilde \gamma=\widetilde \gamma_k$.
We first continue the discussion in Section \ref{problem} on the impact of multiple RF sources and multiple receive antennas.
Then, the BER performance of all the proposed detectors will be illustrated.
Finally, the effectiveness of the proposed channel estimation method will be evaluated.

\subsection{Achievable Rate and CI for Different $K$ and $M$}\label{OOK_rate_sim}

\begin{figure}
[!t]
\centering
\includegraphics[width=.99\columnwidth]{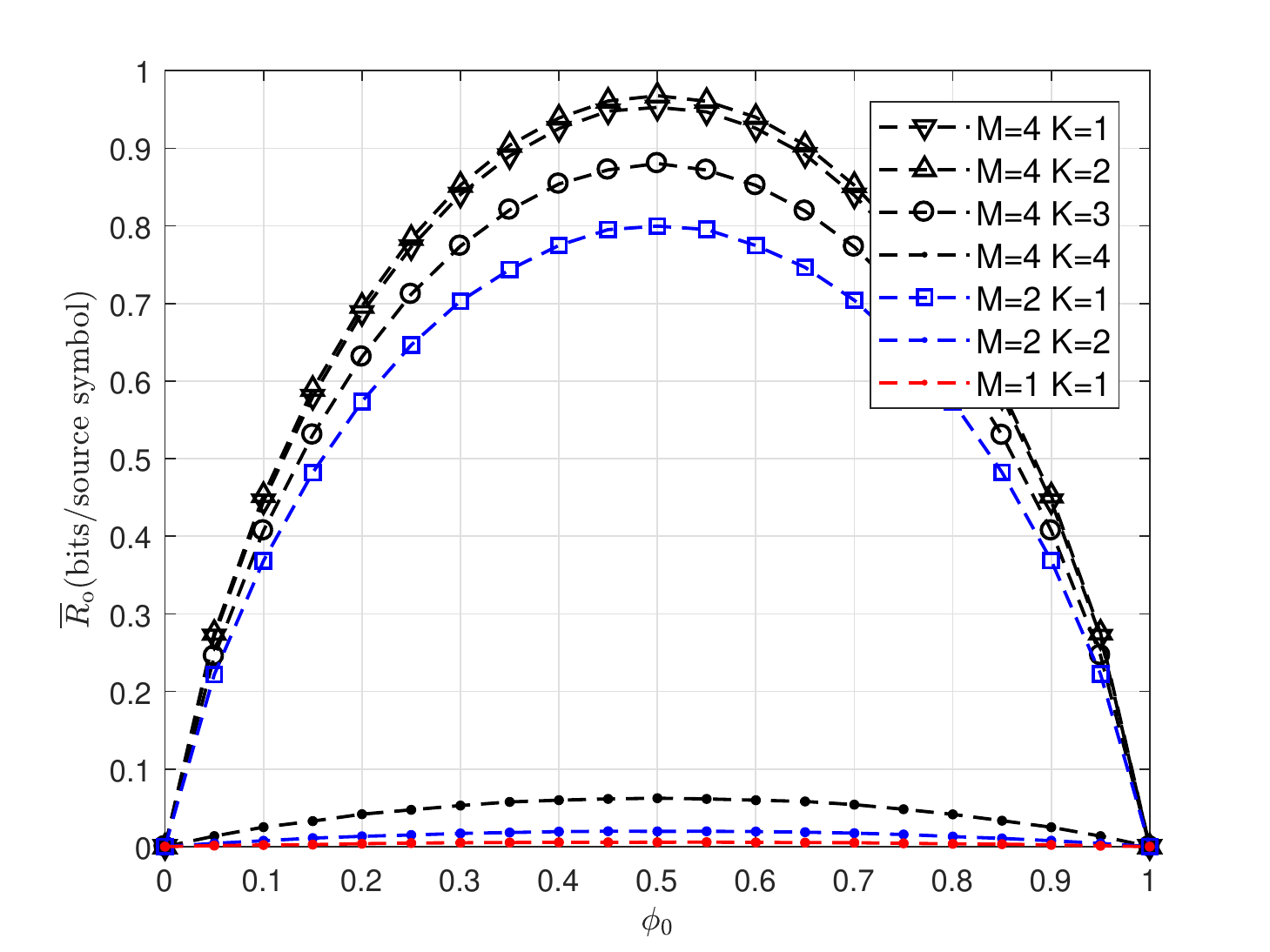}
\caption{The average throughput versus $\phi_0$ (the prior probability that $c=0$) when the BD adopts OOK, $\gamma_{\rm d}=40$ dB, and $\widetilde \gamma=-25$ dB.
}
\label{ergodic_rate}
\end{figure}

{\figurename~\ref{ergodic_rate}} illustrates the relationship between the average throughput $\overline{R}_{\rm o}$ and the system parameters, such as the prior probability of transmit symbol ``0'' (i.e., $\phi_0$), antenna number $M$, and amount of RF sources $K$.
The average throughput $\overline{R}_{\rm o}$ is obtained through  Monte Carlo integration according to Appendix \ref{A_onoff_rate}.
It is noticed that, for all the combination cases of $M$ and $K$ considered in {\figurename~\ref{backscatter model}}, $\phi_0=0.5$ always achieves the maximum average throughput. Thus to transmit equiprobable symbols is optimal.
In addition, one can also see the significant improvement of $\overline{R}_{\rm o}$ when the condition $M>K$ is satisfied.
For the $M=K$ cases, the average throughput still increases slightly when $M$ becomes larger.
In the rest numerical discussions, we set BD transmits ``0'' and ``1'' with equal probability.

\begin{figure}
[!t]
\centering
\includegraphics[width=.99\columnwidth]{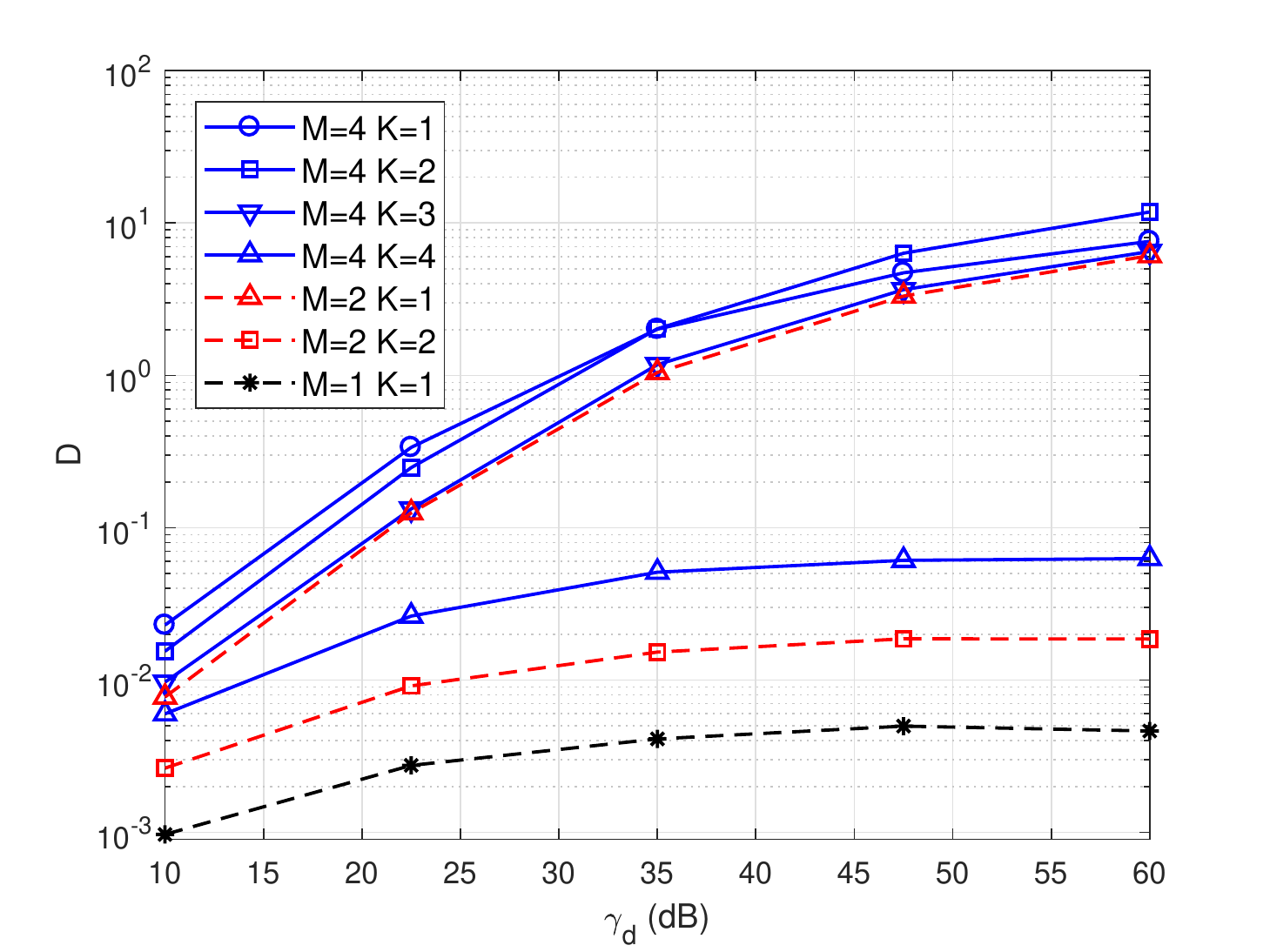}
\caption{The CI versus $\gamma_{\rm d}$ when the BD transmits ``0'' and ``1'' with equal probability, and $\widetilde \gamma=-25$ dB.
}
\label{CI_vs_rd}
\end{figure}

{\figurename~\ref{CI_vs_rd}} plots the CI (i.e., $D$ in the $Y$-axis) for different combinations of $M$ and $K$ with respect to the direct link SNR $\gamma_{\rm{d}}$ when $\widetilde \gamma=-25$ dB, when BD employs on-off modulation.
It is seen that, if $M>K$ is satisfied, $D$ increases as the $\gamma_{\rm{d}}$ increases.
However, when $M=K$, $D$ approaches to an finite upper bound with the increase of $\gamma_{\rm{d}}$.
Therefore, when $\gamma_{\rm{d}}$ is already high enough, to increase $\gamma_{\rm{d}}$ contributes little to the detection performance for the $M=K$ cases.
In addition, we observes an interesting phenomenon that, the CI of $\{M=4, K=2\}$ is high than that of $\{M=4, K=1\}$ when $\gamma_{\rm{d}}>35$ dB.
Hence, a larger $K$ does not necessarily implies poorer performance.
This is because that, although the multiple RF sources consume the degree of freedom for DLI cancelation, they also provides more carrier emitter energy which benefits the backscatter transmission.
Furthermore, it is shown that, the CI of $\{M=4, K=3\}$ is slightly better than that of $\{M=2, K=1\}$, although the rate $\frac{K}{M}$ of the $\{M=4, K=3\}$ cases is larger.

\subsection{BER Performance}
In this subsection, we evaluate the effectiveness of the proposed detectors by simulation.
The backscatter frame length is fixed to $I=100$. In addition, $10^8$ Monte-Carlo runs are carried out to achieve reliable results.

\begin{figure}
[!t]
\centering
\includegraphics[width=.99\columnwidth]{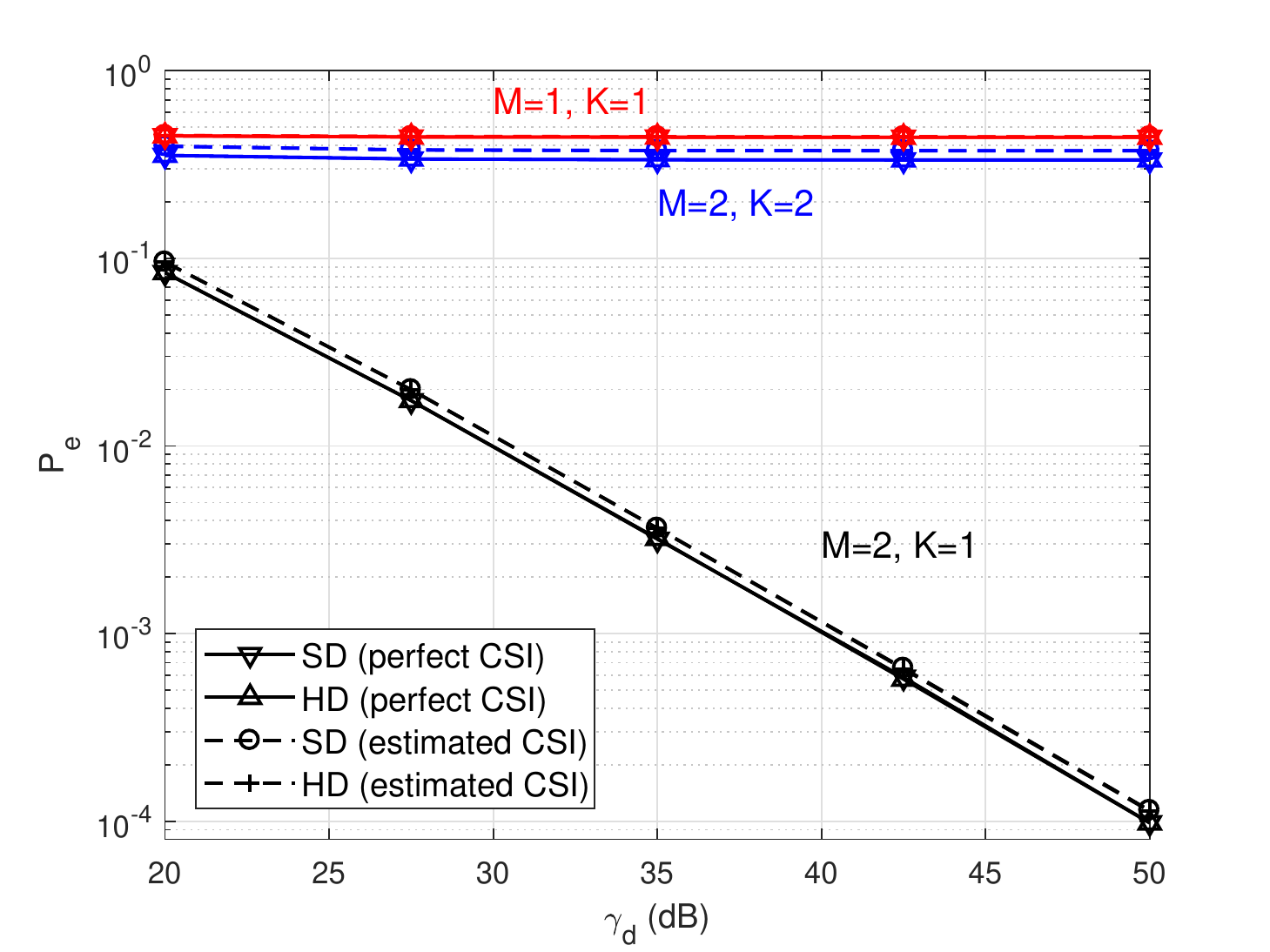}
\caption{The BER versus $\gamma_d$, when $N=50$, $\widetilde \gamma = -25$ dB and $I=100$, for $M=2$ and $M=1$, respectively.
}
\label{BER_M2}
\end{figure}

\begin{figure}
[!t]
\centering
\includegraphics[width=.99\columnwidth]{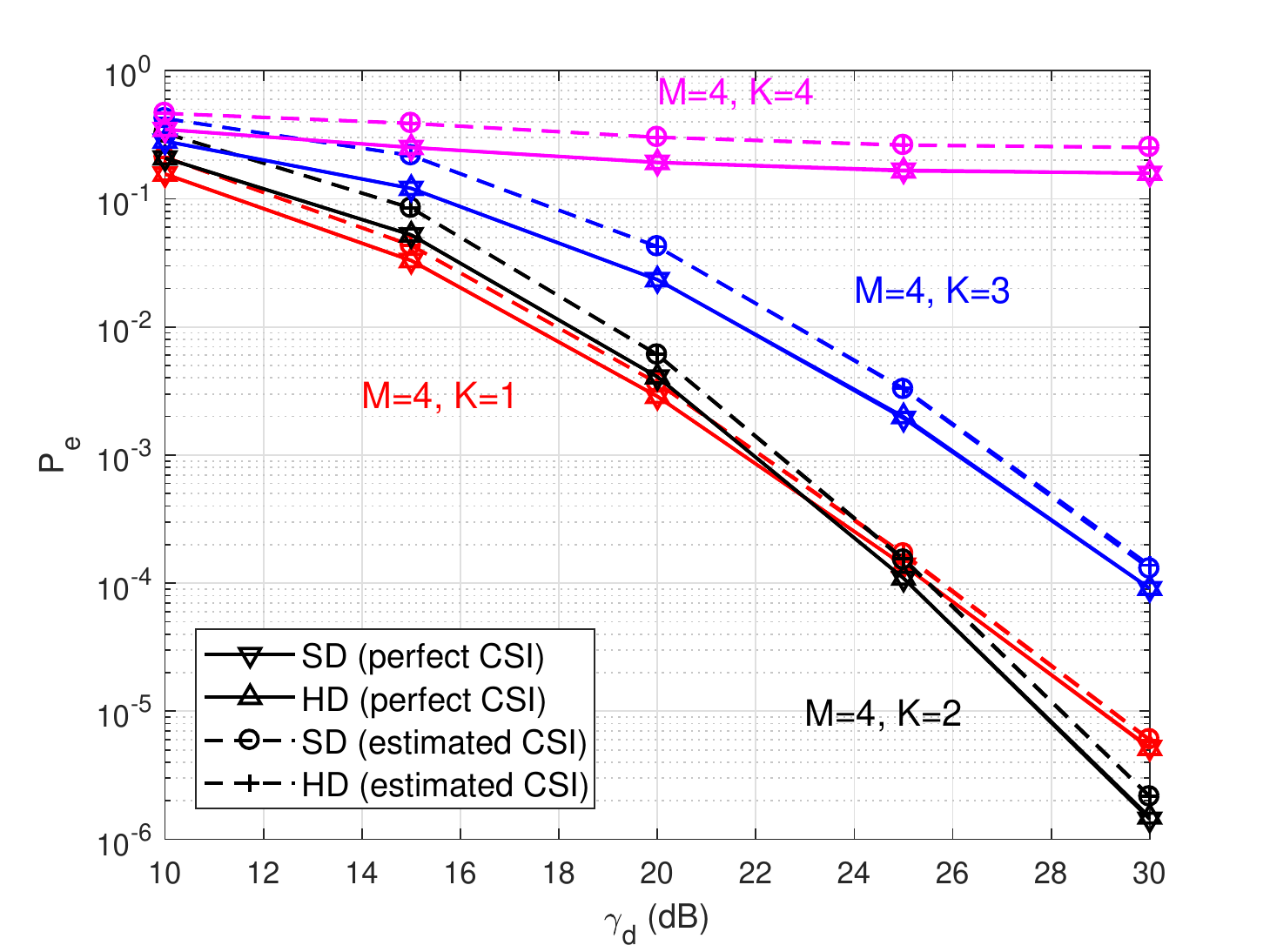}
\caption{The BER versus $\gamma_d$, when $N=50$, $\widetilde \gamma = -25$ dB, $I=100$, and $M=4$, for different $K$.
}
\label{BER_M4}
\end{figure}

{\figurename~\ref{BER_M2}} illustrates the BER performance versus the direct link SNR $\gamma_{\rm{d}}$ for different detectors with $N=50$ and $\Delta \gamma = -25$ dB.
Specifically, the BERs of the optimal detector (i.e., SD) and suboptimal detector (i.e., HD) with perfect and estimated CSI are displayed, respectively.
It is seen that the HD achieves the same BER performance as that of the SD.
Besides, the detectors with estimated CSI suffer only a very small performance loss compared to the perfect CSI cases.
In addition, the BER meet the error floor when $M=K$.
However, when $M=2$ and $K=1$, the BER decreases significantly with the increase of $\gamma_{\rm{d}}$. Therefore, the error floor problem is efficiently eliminated.

{\figurename~\ref{BER_M4}} illustrates the BER performance versus $\gamma_{\rm{d}}$ for different $K$ when $M=4$ and $N=50$, $\Delta \gamma = -25$ dB.
The HD and SD still achieve almost the same performance.
Therefore, the performance loss caused by channel estimation is very small.
It is noticed that the error floor problem is existed when $K=4$, and it is overcome when $K\leq 3$.
Furthermore, when $K=1$ and $K=2$, the detectors achieve similar performance.
When $\gamma_{\rm{d}}$ increases, the BER performance of $K=2$ becomes better than that of $K=1$ cases, which is consistent with the prediction from {\figurename~\ref{CI_vs_rd}}.

\begin{figure}
[!t]
\centering
\includegraphics[width=.99\columnwidth]{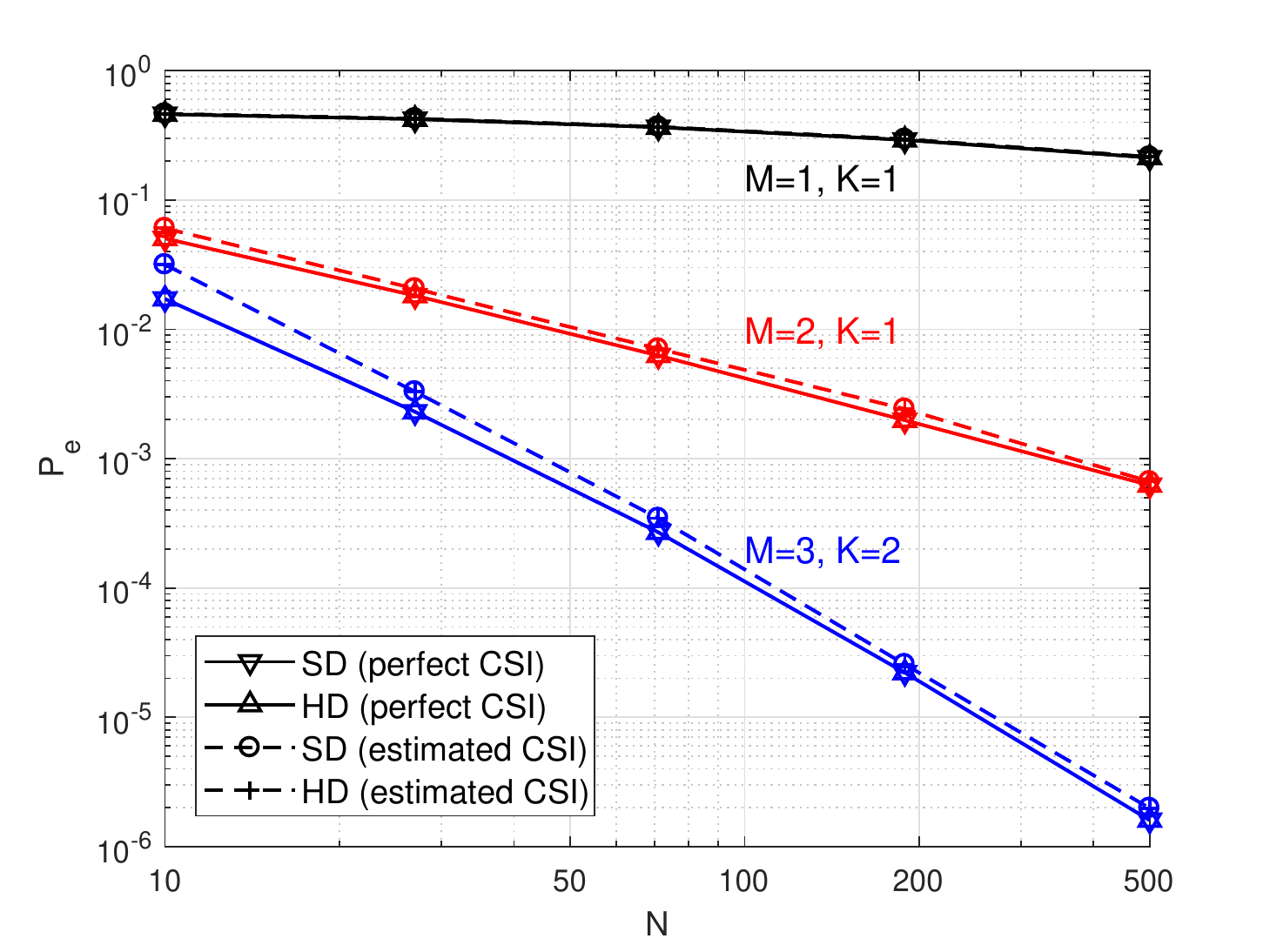}
\caption{The BER versus $N$, when $\gamma_{\rm{d}}=25$ dB, $\widetilde \gamma = -20$ dB, and $I=100$ for different $K$ and $M$.
}
\label{BER_N}
\end{figure}

{\figurename~\ref{BER_N}} plots the BER performance of the proposed detectors for different $N$, when $\gamma_{\rm{d}}=25$ dB and $\Delta \gamma = -20$ dB.
It is noticed that the performance of all detectors improves with the increase of $N$ at the cost of lower transmission rate.
For a target BER of $10^{-2}$, when $M=3$ and $K=2$, each backscatter symbol needs to span less than $N=20$ RF source symbols.
When $M=2$ and $K=1$, the backscatter symbol needs to span about $N=50$ RF source symbols.
However, when $M=1$ and $K=1$, the BER is unacceptable even when $N=500$.
Therefore, the transmission rate can be improved at least $10$ times, when multiple antennas is exploited and $M>K$ is satisfied.
In addition, we observe that the BER performance of $\{M=3,K=2\}$ is better than that of $\{M=2,K=1\}$, although the rate $\frac{K}{M}$ is increased.

\subsection{Channel Estimation Error}
Finally, the quality of the channel estimation by the proposed clustering-based method are assessed in this subsection.
The \emph{normalized mean squared error} (NMSE) is chosen as the evaluation metric, which is defined as follows\footnote{It should be noted that, the proposed clustering method only needs to classify the received samples into two groups, and it cannot (actually does not need to) map each group to $c=0$ and $c=1$, respectively. Thus, to apply \eqref{equ:NMSE}, we need to check the transmitted $c^{(a_0)}$ and $c^{(a_1)}$ in simulations to realize mapping.}:
\begin{equation}\label{equ:NMSE}
J_{\rm{NMSE}}=  {\mathbb E} \left[ \frac{1}{2} \sum_{i=0}^{1} \frac {\| {\bf{C}}_i - \hat{\bf{C}}_i \|_{\rm F}^2 }
        {\| {\bf{C}}_i \|_{\rm F}^2 }
  \right]
.
\end{equation}
We suppose that the receiver equips $M=4$ antennas, and the backscatter frame consists of $I=100$ backscatter symbols.

\begin{figure}
[!t]
\centering
\includegraphics[width=.99\columnwidth]{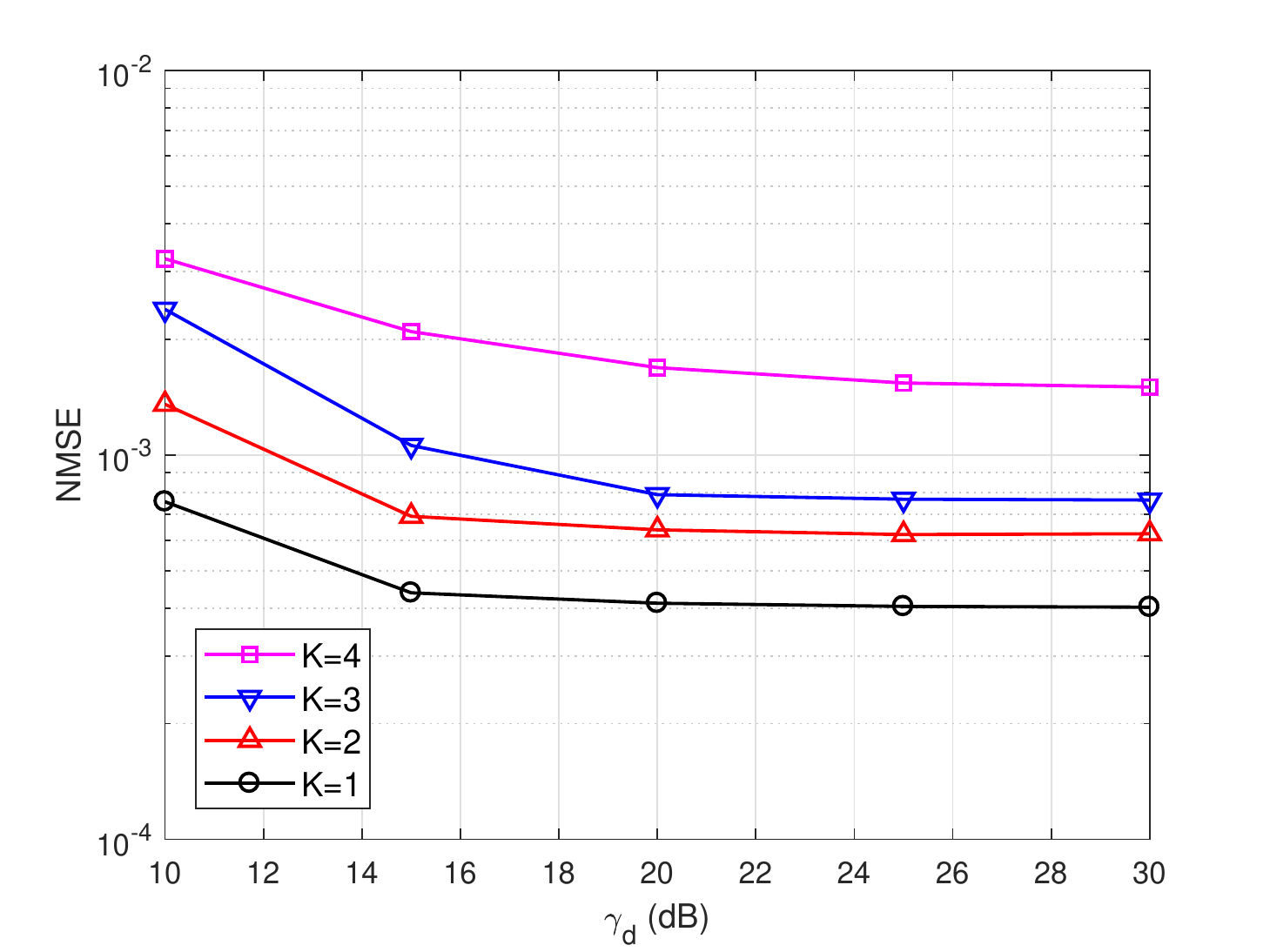}
\caption{The NMSE versus $\gamma_d$, when $\widetilde \gamma = -20$ dB, $N=50$, $I=100$, and $M=4$, for different $K$.
}
\label{NMSE_rd}
\end{figure}

{\figurename~\ref{NMSE_rd}} shows the NMSE performance over a range of $\gamma_d$ when $K$ increases from $1$ to $4$, in which $\widetilde \gamma = -20$ dB and $N=50$.
It is seen that the NMSE converges to a constant with the increase of $\gamma_d$.
This implies that, although the final detection BER can be improved as $\gamma_d$ increases when $M>K$, the channel estimation error always exists given a finite $\widetilde \gamma$ and $N$.
In addition, it is noticed that, when $K$ increases, the NMSE increases, since the number of directions of arrival required to be estimated increases in proportion to $K$.

%

\begin{figure}
[!t]
\centering
\includegraphics[width=.99\columnwidth]{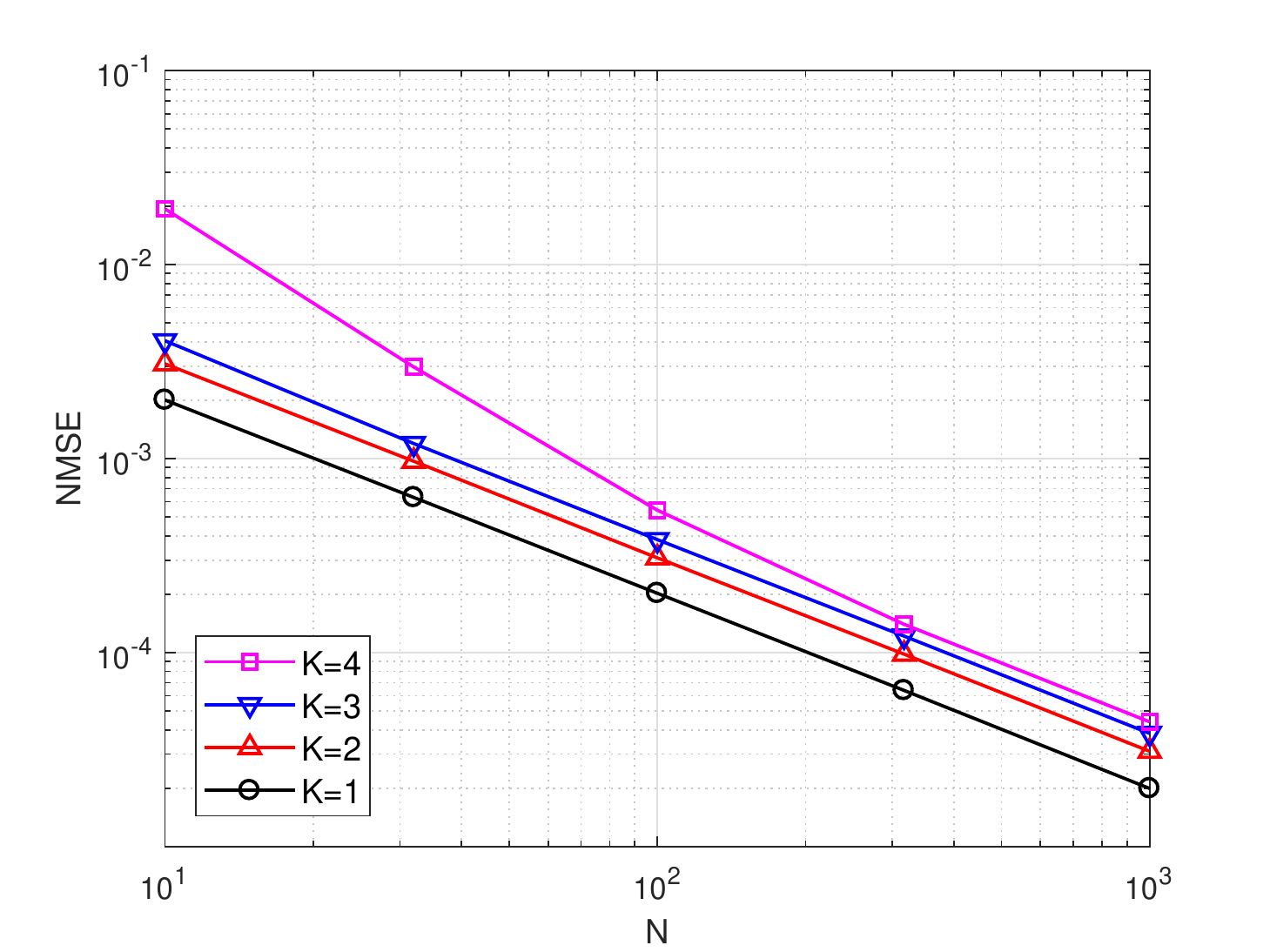}
\caption{The NMSE versus $N$, when $\gamma_d = 30$ dB, $\widetilde \gamma=-20$ dB, $I=100$, and $M=4$, for different $K$.
}
\label{NMSE_N}
\end{figure}

{\figurename~\ref{NMSE_N}} illustrate the NMSE with regard to the spreading factor $N$.
The parameters are set as $\gamma_d = 30$ dB and $\widetilde \gamma=-20$ dB.
It is seen that the NMSE for all the chosen $K$ significantly decreases with the increase of $N$.
The reason is that, the quality of the sample covariance matrix ${{\bf{R}}}^{(i)}$ in \eqref{equ:samplecm} becomes higher as $N$ increases, which results in better estimation performance.

\section{Conclusion}\label{conclusion}
In this paper, we have investigated on the cognitive backscatter signal transmission, where multiple RF sources are considered, and there is no cooperative between the legacy system and the backscatter system.
Our study has demonstrated that to exploit multiple antennas at the receiver is a feasible way to achieve high-throughput backscatter signal transmission.
The optimal SD and suboptimal HD detectors have been designed based on the ML criterion.
In addition, a fully blind channel estimation method has been proposed based on the statistical clustering.
Simulation results have been provided to evaluated the performance of the proposed detectors.
It is verified that, the error-floor phenomenon can be avoided when $M>K$,  the HD based suboptimal detector can achieve almost the optimal performance with lower complexity, and the detector using the estimated CSI can perform comparably as their counterparts with perfect CSI.

\appendices
\section{Proof of Proposition \ref{pro2}}\label{M_larger_K}
One simple method to suppress the DLIs from $K$ RF sources is the the \emph{zero-forcing} (ZF) beamforming receiver \cite{Tse2005Wirelesscom}, which projects the received signal ${\bf{y}}$ onto the subspace orthogonal to the one spanned by the vectors ${{\bf{f}}_1},{{\bf{f}}_2},\cdots,{{\bf{f}}_K}$\footnote{It is noted that ZF beamforming receiver is a suboptimal but simple linear receiver. We choose it in this proof to provide a performance lower bound.
In practice, the \emph{minimum mean square error} (MMSE) receiver or the ML receiver can be implemented to achieve better performance.}.

Letting ${\bf H}=[{{\bf{f}}_1},{{\bf{f}}_2},\cdots,{{\bf{f}}_K}, {\bf{g}}]$, the ZF beamforming vector ${\bf w}$ is the last row of the pseudoinverse ${\bf H}^\dagger$ of the matrix ${\bf H}$, defined by:
\begin{equation}
{\bf H}^\dagger := ({\bf H}^H {\bf H})^{-1} {\bf H}^H.
\end{equation}
Since $K<M$ is satisfied, we have ${\bf{w}} {{\bf{f}}_k}=0$ for $k=1,\cdots,K$ \cite{Tse2005Wirelesscom}.
Using ${\bf{w}}_{\rm{D}}=\frac{{\bf{w}}}{\|{\bf{w}}\|}$, the interference in \eqref{equ:Mantenna} is removed:
\begin{equation}\label{equ:ZF_cancel}
\begin{aligned} [b]
x&= {\bf{w}}_{\rm{D}} {\bf{y}}\\
&={\bf{w}}_{\rm{D}} {\bf{g}} \sum_{k=1}^K l_k \sqrt{\widetilde \gamma_k {\gamma_{{\rm{d}},k}}} s_k c+{\bf{w}}_{\rm{D}} {\bf{u}}
.
\end{aligned}
\end{equation}

Let ${\bf s}=[s_1,s_2,\cdots,s_K]$. From \eqref{equ:ZF_cancel}, ${\bf s}$  can be combined into the channel response. Then the achievable rate of the backscatter signal using ZF beamforming receiver is
\begin{equation}\label{equ:R_ZF}
\begin{aligned} [b]
{\widetilde{R}}_{\{M,K\}} &=  {\mathbb E}_{\{ {\bf{s}} \}} \left[ \log_2 (1+
    \frac   {\left|{\bf{w}}_{\rm{D}} {\bf{g}} \sum_{k=1}^K l_k \sqrt{\widetilde \gamma_k {\gamma_{{\rm{d}},k}}} s_k\right|^2}
            { \| {\bf{w}}_{\rm{D}} \|^2}
) \right]\\
&=  {\mathbb E}_{\{ {\bf{s}} \}} \left[ \log_2 (1+
       {\left|{\bf{w}}_{\rm{D}} {\bf{g}}\right|^2 \sum_{k=1}^K {\widetilde \gamma_k} {\gamma_{{\rm{d}},k}} |l_k s_k|^2}
) \right]
.
\end{aligned}
\end{equation}
Obviously, we have ${{R}}_{\{M,K\}}\geq {\widetilde{R}}_{\{M,K\}}$.

Assuming ${\gamma_{{\rm{d}},k_0}}$ is infinite, we have
\begin{equation}\label{equ:R_ZF_k0}
\begin{aligned} [b]
{\widetilde{R}}_{\{M,K\}} &\geq  {\mathbb E}_{\bf s} \left[ \log_2 (1+
       {{\gamma_{{\rm{d}},k_0}} {\widetilde \gamma_{k_0}}  \left|{\bf{w}}_{\rm{D}} {\bf{g}}\right|^2 |l_{k_0}|^2 |s_{k_0}|^2}
) \right]
.
\end{aligned}
\end{equation}
Since $s_{k_0} \sim{\cal{CN}}(0,1)$, $|s_{k_0}|^2$ follows exponential distribution.
By letting $\delta={{\gamma_{{\rm{d}},k_0}} {\widetilde \gamma_{k_0}}  \left|{\bf{w}}_{\rm{D}} {\bf{g}}\right|^2 |l_{k_0}|^2}$, \eqref{equ:R_ZF_k0} becomes
\begin{equation}\label{equ:R_ZF_k0a}
\begin{aligned} [b]
{\widetilde{R}}_{\{M,K\}} &\geq  {\mathbb E}_{\{ {\bf{s}} \}} \left[ \log_2 (1+
       {\delta |s_{k_0}|^2}
) \right]\\
&=\int_0^{+\infty} { \log_2(1+\delta x) e^{-x}
   }{\rm d}x\\
&=\frac{e^{\frac{1}{\delta}}}{\ln 2} \int_{\frac{1}{\delta}}^{+\infty} { \frac {e^{-x}}{x}
   }{\rm d}x
.
\end{aligned}
\end{equation}
Finally, we have
\begin{equation} \label{equ:Rdown_Mantenna}
\begin{aligned} [b]
R_{\{M,K\}}^{\infty}&= \lim_{{\gamma_{{\rm{d}},k_0}} \rightarrow +\infty} R_{\{M,K\}}\\
&\geq \lim_{{\gamma_{{\rm{d}},k_0}} \rightarrow +\infty} {\widetilde{R}}_{\{M,K\}}\\
&\geq \lim_{\delta \rightarrow +\infty}  \frac{e^{\frac{1}{\delta}}}{\ln 2} \int_{\frac{1}{\delta}}^{+\infty} { \frac {e^{-x}}{x}   }{\rm d}x\\
&=\frac{1}{\ln 2} \int_0^{+\infty} { \frac {e^{-x}}{x}   }{\rm d}x\\
&=+\infty
.
\end{aligned}
\end{equation}

\section{The Maximum Rate of the On-off Modulated $c$}\label{A_onoff_rate}
Let $R_{\rm o}$ denote the achievable rate for the on-off modulated $c$, which is the mutual information between the OOK modulated $c$ and received signal ${\bf{y}}$, i.e., $R_{\rm o} =  {\rm I}(c;{\bf{y}})$.
Then the average maximum  achievable rate $\overline{R}_{\rm o}$ is:
\begin{equation}\label{equ:ER_ook}
\overline{R}_{\rm o} = {\mathbb E}_{\{  {\bf{g}}, {\bf{f}},l \}} \left[ {\rm I}(c;{\bf{y}})\right]
.
\end{equation}
It is known that, the close-form expression of $R_{\rm o}$ is unavailable even for the $M=1$ cases \cite{WangG2018AmBCrate}.
Thus in this section, we try to present a numerical method to obtain $\overline{R}_{\rm o}$.

The mutual information ${\rm I}(c;{\bf{y}})$ is expressed as follows
\begin{equation}\label{equ:R_ookn}
\begin{aligned} [b]
{\rm I}(c;{\bf{y}})
&={\rm H}(c)-{\rm H}(c|{\bf{y}})\\
&={\rm H_b}(\phi_0)+{\mathbb E}_{\left\{{\bf{y}}_0\right\}} \left[ {\rm H}(c|{\bf{y}}_0) \right]
.
\end{aligned}
\end{equation}
where ${\bf{y}}_0$ is once realization of ${\bf{y}}$, $\phi_0$ and $\phi_1$ are the prior probabilities when ``$c=0$'' and ``$c=1$'', respectively, which satisfies $\phi_0+\phi_1=1$, ${\rm H_b}(\phi_0)\triangleq -\phi_0 \log_2 \phi_0-\phi_1 \log_2 \phi_1$ is the binary entropy function, and $p( c | {\bf{y}}_0)$ is the posterior probability of $c$ while receiving ${\bf{y}}_0$.

Since ${\rm H_b}(\phi_0)$ is independent of all the channel coefficients, the average achievable rate has the same expression as \eqref{equ:R_ookn}:
\begin{equation}\label{equ:ER_ookn}
\begin{aligned} [b]
\overline{R}_{\rm o} &= {\mathbb E}_{\{ {\bf{g}}, {\bf{f}},l \}} {\rm I}(c;{\bf{y}})\\
&={\rm H_b}(\phi_0)+{\mathbb E}_{\left\{ {\bf{y}}_0\right\}} \left[ {\rm H}(c|{\bf{y}}_0)\right]
.
\end{aligned}
\end{equation}
Note that, in \eqref{equ:R_ookn} and \eqref{equ:ER_ookn}, ${\rm H}(c|{\bf{y}}_0)$ is averaged with respect to ${\bf{y}}_0$ according to different prior distributions.

Using the PDFs in \eqref{equ:Gpdfc1} and \eqref{equ:Gpdfc0}, we have the posterior probability as follows:
\begin{equation}\label{equ:post_y}
\begin{aligned} [b]
p( c=j | {\bf{y}}_0) &= \frac{\phi_j p( {\bf{y}}_0| c=j) } {p({\bf{y}}_0)}
,
\end{aligned}
\end{equation}
where $p({\bf{y}}_0) = \phi_0 {p( {\bf{y}}_0| c=0) } +\phi_1 {p( {\bf{y}}_0| c=1) }$.
Let $\varepsilon_j=p( c=j | {\bf{y}}_0)$, for $j=0,1$.
Then we have the conditional entropy
\begin{equation}\label{equ:conditionalE}
\begin{aligned} [b]
{\rm H}(c|{\bf{y}}_0)&=- \sum_{j=0}^1 \varepsilon_j \log_2 \varepsilon_j \\
&={\rm H_b}(\varepsilon_0).
\end{aligned}
\end{equation}

Substituting \eqref{equ:conditionalE} into \eqref{equ:ER_ookn}, the average achievable rate is
\begin{equation}\label{equ:R_ookF}
\begin{aligned} [b]
\overline{R}_{\rm o}
&= {\rm H_b}(\phi_0)-{\mathbb E}_{\left\{ {\bf{y}}_0\right\}} \left[{\rm H_b}(\varepsilon)\right]\\
&= {\rm H_b}(\phi_0)-\int_{{\bf{y}}_0} {p({\bf{y}}_0) {\rm H_b}(\varepsilon) } {\rm d}{{\bf{y}}_0}.
\end{aligned}
\end{equation}
The integral in \eqref{equ:R_ookF} can be calculated by Monte Carlo integration. Firstly, the channel coefficients ${\bf{g}}$, ${\bf{f}}_1,\cdots,{\bf{f}}_K$, and $l_1,\cdots,l_K$ are randomly generalized, and then  ${\bf{C}}_0$ and ${\bf{C}}_1$ are obtained for each realization based on \eqref{equ:C0} and \eqref{equ:C1}.
Secondly, $c$ and ${\bf{y}}_0$ are randomly generated for sufficient times given a group of $\phi_0$, $\phi_1$, ${\bf{C}}_0$ and ${\bf{C}}_1$,
and then an instantaneous result $R_t={\rm H_b}(\phi_0)-{ {\rm H_b}(\varepsilon) }$ is obtained.
Finally, the mean value of $R_t$ will well approximate the $\overline{R}_{\rm o}$ in \eqref{equ:R_ookF} based on the \emph{Law of Large Numbers}.

%
%
%
%
%
%
%
%

%
%
%
%
%
%
%

\ifCLASSOPTIONcaptionsoff
  \newpage
\fi

%
%
%
%
%

\bibliographystyle{IEEEtran}
\bibliography{IEEEabrv,mybib_draft2}

\begin{thebibliography}{10}
\providecommand{\url}[1]{#1}
\csname url@samestyle\endcsname
\providecommand{\newblock}{\relax}
\providecommand{\bibinfo}[2]{#2}
\providecommand{\BIBentrySTDinterwordspacing}{\spaceskip=0pt\relax}
\providecommand{\BIBentryALTinterwordstretchfactor}{4}
\providecommand{\BIBentryALTinterwordspacing}{\spaceskip=\fontdimen2\font plus
\BIBentryALTinterwordstretchfactor\fontdimen3\font minus
  \fontdimen4\font\relax}
\providecommand{\BIBforeignlanguage}[2]{{%
\expandafter\ifx\csname l@#1\endcsname\relax
\typeout{** WARNING: IEEEtran.bst: No hyphenation pattern has been}%
\typeout{** loaded for the language `#1'. Using the pattern for}%
\typeout{** the default language instead.}%
\else
\language=\csname l@#1\endcsname
\fi
#2}}
\providecommand{\BIBdecl}{\relax}
\BIBdecl

\bibitem{Stankovic2014DirectionIoT}
J.~A. Stankovic, ``Research directions for the internet of things,'' \emph{IEEE
  Internet Things J.}, vol.~1, no.~1, pp. 3--9, 2014.

\bibitem{LIU2013AMBC}
V.~Liu, A.~Parks, V.~Talla, S.~Gollakota, D.~Wetherall, and J.~R. Smith,
  ``Ambient backscatter: wireless communication out of thin air,'' in
  \emph{Proc. ACM SIGCOMM}, Aug 2013, pp. 39--50.

\bibitem{Welbourne2009RFID}
E.~Welbourne, L.~Battle, G.~Cole, K.~Gould, K.~Rector, S.~Raymer,
  M.~Balazinska, and G.~Borriello, ``Building the {I}nternet of {T}hings using
  {RFID}: The {RFID} ecosystem experience,'' \emph{IEEE Internet Comput.},
  vol.~13, no.~3, pp. 48--55, 2009.

\bibitem{Boyer2014RFIDBC}
C.~Boyer and S.~Roy, ``Backscatter communication and {RFID}: Coding, energy,
  and {MIMO} analysis,'' \emph{IEEE Trans. Commun.}, vol.~62, no.~3, pp.
  770--785, 2014.

\bibitem{ZhangR2013WIPT}
L.~Liu, R.~Zhang, and K.~C. Chua, ``Wireless information and power transfer: A
  dynamic power splitting approach,'' \emph{IEEE Trans. Commun.}, vol.~61,
  no.~9, pp. 3990--4001, 2013.

\bibitem{Larsson2013SIPT}
K.~Huang and E.~Larsson, ``Simultaneous information and power transfer for
  broadband wireless systems,'' \emph{IEEE Trans. Signal Process.}, vol.~61,
  no.~23, pp. 5972--5986, 2013.

\bibitem{KangX2017RidingPrimary}
X.~Kang, Y.~C. Liang, and J.~Yang, ``Riding on the primary: A new spectrum
  sharing paradigm for wireless-powered {I}o{T} devices,'' in \emph{Proc. IEEE
  ICC}, May 2017, pp. 1--6.

\bibitem{Niyato2018survey}
N.~V. Huynh, D.~T. Hoang, X.~Lu, D.~Niyato, P.~Wang, and D.~I. Kim, ``Ambient
  backscatter communications: A contemporary survey,'' \emph{Commun. Surveys
  Tuts.}, Early Access.

\bibitem{Niyato2018AmBCMagazine}
X.~Lu, D.~Niyato, H.~Jiang, D.~I. Kim, Y.~Xiao, and Z.~Han, ``Ambient
  backscatter assisted wireless powered communications,'' \emph{IEEE Wireless
  Commun.}, vol.~25, no.~2, pp. 170--177, 2018.

\bibitem{LiuW2017survey}
W.~{Liu}, K.~{Huang}, X.~{Zhou}, and S.~{Durrani}, ``{Next Generation
  Backscatter Communication: Theory and Applications},'' \emph{ArXiv e-prints},
  Jan. 2017.

\bibitem{Darsena2017ModelandPerformance}
D.~Darsena, G.~Gelli, and F.~Verde, ``Modeling and performance analysis of
  wireless networks with ambient backscatter devices,'' \emph{IEEE Trans.
  Commun.}, vol.~65, no.~4, pp. 1797--1814, 2017.

\bibitem{darsena2016performance}
------, ``Performance analysis of ambient backscattering for green internet of
  things,'' in \emph{IEEE PIMRC}, 2016, pp. 1--6.

\bibitem{WangG2018AmBCrate}
W.~Zhao, G.~Wang, R.~Fan, L.~S. Fan, and S.~Atapattu, ``Ambient backscatter
  communication systems: Capacity and outage performance analysis,'' \emph{IEEE
  Access}, vol.~6, pp. 22\,695--22\,704, 2018.

\bibitem{Dinesh2015BackFi}
D.~Bharadiay, K.~Joshiy, M.~Kotaru, and S.~Katti, ``Back{F}i: {H}igh throughput
  {W}i{F}i backscatter,'' \emph{ACM SIGCOMM}, vol.~45, no.~4, pp. 283--296,
  2015.

\bibitem{Ensworth2017BLE}
J.~F. Ensworth and M.~S. Reynolds, ``{BLE}-backscatter: Ultralow-power {IoT}
  nodes compatible with bluetooth 4.0 low energy ({BLE}) smartphones and
  tablets,'' \emph{IEEE Trans. Microw. Theory Techn.}, vol.~65, no.~9, pp.
  3360--3368, 2017.

\bibitem{Talla2017LoRabackscatter}
V.~Talla, M.~Hessar, B.~Kellogg, A.~Najafi, J.~R. Smith, and S.~Gollakota,
  ``Lora backscatter: Enabling the vision of ubiquitous connectivity,''
  \emph{Proc. ACM Interact. Mob. Wearable Ubiquitous Technol.}, vol.~1, no.~3,
  p. 105, 2017.

\bibitem{lyu2018optimal}
B.~Lyu, C.~You, Z.~Yang, and G.~Gui, ``The optimal control policy for
  {RF}-powered backscatter communication networks,'' \emph{IEEE Trans. Veh.
  Technol.}, vol.~67, no.~3, pp. 2804--2808, 2018.

\bibitem{Niyato2017ImproveNetPerform}
D.~T. Hoang, D.~Niyato, P.~Wang, D.~I. Kim, and Z.~Han, ``Ambient backscatter:
  A new approach to improve network performance for {RF}-powered cognitive
  radio networks,'' \emph{IEEE Trans. Commun.}, vol.~65, no.~9, pp. 3659--3674,
  2017.

\bibitem{Niyato2018TimeSchedulingHTB}
N.~V. Huynh, D.~T. Hoang, D.~Niyato, P.~Wang, and D.~I. Kim, ``Optimal time
  scheduling for wireless-powered backscatter communication networks,''
  \emph{IEEE Wireless Commun. Lett.}, EArly Access.

\bibitem{wang2018stackelberg}
W.~Wang, D.~T. Hoang, D.~Niyato, P.~Wang, and D.~I. Kim, ``Stackelberg game for
  distributed time scheduling in {RF}-powered backscatter cognitive radio
  networks,'' \emph{IEEE Trans. Wireless Commun.}, Early Access.

\bibitem{lu2018wireless}
X.~Lu, H.~Jiang, D.~Niyato, D.~I. Kim, and Z.~Han, ``Wireless-powered
  device-to-device communications with ambient backscattering: Performance
  modeling and analysis,'' \emph{IEEE Trans. Wireless Commun.}, vol.~17, no.~3,
  pp. 1528--1544, 2018.

\bibitem{TaoQ2018Manchester}
Q.~Tao, C.~Zhong, H.~Lin, and Z.~Zhang, ``Symbol detection of ambient
  backscatter systems with manchester coding,'' \emph{IEEE Trans. Wireless
  Commun.}, vol.~17, no.~6, pp. 4028--4038, 2018.

\bibitem{GaoFF2017EDAmBC}
J.~Qian, F.~Gao, G.~Wang, S.~Jin, and H.~Zhu, ``Semi-coherent detection and
  performance analysis for ambient backscatter system,'' \emph{IEEE Trans.
  Commun.}, vol.~65, no.~12, pp. 5266--5279, 2017.

\bibitem{wang2016ambient}
G.~Wang, F.~Gao, R.~Fan, and C.~Tellambura, ``Ambient backscatter communication
  systems: Detection and performance analysis,'' \emph{IEEE Trans. Commun.},
  vol.~64, no.~11, pp. 4836--4846, 2016.

\bibitem{GaoFF2017NDAmBC}
J.~Qian, F.~Gao, G.~Wang, S.~Jin, and H.~Zhu, ``Noncoherent detections for
  ambient backscatter system,'' \emph{IEEE Trans. Wireless Commun.}, vol.~16,
  no.~3, pp. 1412--1422, 2017.

\bibitem{Guo2018IoT}
H.~Guo, Q.~Zhang, S.~Xiao, and Y.~C. Liang, ``Exploiting multiple antennas for
  cognitive ambient backscatter communication,'' \emph{IEEE Internet Things
  J.}, Early Access.

\bibitem{YangG2017CooperativeBC}
G.~Yang, Q.~Zhang, and Y.~C. Liang, ``Cooperative ambient backscatter
  communications for green {I}nternet-of-{T}hings,'' \emph{IEEE Internet Things
  J.}, vol.~5, no.~2, pp. 1116--1130, 2018.

\bibitem{RZLong2017beamform}
R.~Long, G.~Yang, Y.~Pei, and R.~Zhang, ``Transmit beamforming for cooperative
  ambient backscatter communication systems,'' in \emph{IEEE GLOBECOM}, 2017,
  pp. 1--6.

\bibitem{DuanRF2017TVTRbistatic}
R.~Duan, R.~Jantti, H.~Yigitler, and K.~Ruttik, ``On the achievable rate of
  bistatic modulated rescatter systems,'' \emph{IEEE Trans. Veh. Technol.},
  vol.~66, no.~10, pp. 9609--9613, 2017.

\bibitem{YangG2017OFDM}
G.~Yang, Y.~C. Liang, R.~Zhang, and Y.~Pei, ``Modulation in the air:
  Backscatter communication over ambient {OFDM} carrier,'' \emph{IEEE Trans.
  Commun.}, vol.~66, no.~3, pp. 1219--1233, 2018.

\bibitem{cover2012elements}
T.~M. Cover and J.~A. Thomas, \emph{Elements of information theory}.\hskip 1em
  plus 0.5em minus 0.4em\relax John Wiley \& Sons, 2012.

\bibitem{Rappaport2002WCPP}
T.~S. Rappaport, \emph{Wireless communications principles and practice},
  2nd~ed.\hskip 1em plus 0.5em minus 0.4em\relax Prentice Hall, 2002.

\bibitem{Fuschini2008BackscatterAnalyze}
F.~Fuschini, C.~Piersanti, F.~Paolazzi, and G.~Falciasecca, ``Analytical
  approach to the backscattering from {UHF RFID} transponder,'' \emph{IEEE
  Antennas Wireless Propag. Lett.}, vol.~7, pp. 33--35, 2008.

\bibitem{Griffin2009LinkBudget}
J.~D. Griffin and G.~D. Durgin, ``Complete link budgets for backscatter-radio
  and {RFID} systems,'' \emph{IEEE Antennas Propag. Mag.}, vol.~51, no.~2, pp.
  11--25, 2009.

\bibitem{Hastie2009slearn}
T.~Hastie, R.~Tibshirani, and J.~Friedman, \emph{The elements of statistical
  learning: {D}ata mining, inference, and prediction}, 2nd~ed., ser. Springer
  Series in Statistics.\hskip 1em plus 0.5em minus 0.4em\relax Springer, 2009.

\bibitem{Tse2005Wirelesscom}
T.~David and P.~Viswanath, \emph{Fundamentals of Wireless Communication}.\hskip
  1em plus 0.5em minus 0.4em\relax Cambridge University Press, 2005.

\end{thebibliography}


\end{document}